\documentclass[journal,comsoc]{IEEEtran}

\usepackage{amsmath}
\usepackage{amsfonts}
\usepackage{amssymb}
\usepackage[ruled,vlined,boxed,commentsnumbered]{algorithm2e}
\usepackage{setspace}

\usepackage{verbatim}
\usepackage{hyperref}
\usepackage{multirow}
\usepackage{framed}

\usepackage{graphicx}
\usepackage{epstopdf}
\graphicspath{{fig/}}
\usepackage{calc}
\usepackage{color}
\usepackage[caption=false,font=footnotesize]{subfig}

\newtheorem{theorem}{Theorem}

\newtheorem{remark}{Remark}

\newtheorem{example}{Example}
\newtheorem{definition}{Definition}

\newtheorem{stopping criterion}{stopping criterion}
\newtheorem{Restriction}{Restriction}

\newcommand{\beq}{\begin{equation}}
\newcommand{\eeq}{\end{equation}}
\newcommand{\beqnn}{\begin{equation*}}
\newcommand{\eeqnn}{\end{equation*}}
\newcommand{\beqy}{\begin{eqnarray}}
\newcommand{\eeqy}{\end{eqnarray}}
\newcommand{\beqynn}{\begin{eqnarray*}}
\newcommand{\eeqynn}{\end{eqnarray*}}
\newcommand{\bit}{\begin{itemize}}
\newcommand{\eit}{\end{itemize}}
\newcommand{\ben}{\begin{enumerate}}
\newcommand{\een}{\end{enumerate}}
\newcommand{\bex}{\begin{example}}
\newcommand{\eex}{\end{example}}

\newcommand{\balg}[1]{\begin{algorithm} \caption{#1}}
\newcommand{\ealg}{\end{algorithm}}

\newcommand{\balgc}{\begin{algorithmic}[1]}
\newcommand{\ealgc}{\end{algorithmic}}

\newcommand{\bary}{\begin{array}}
\newcommand{\eary}{\end{array}}
\newcommand{\bmx}{\begin{bmatrix}}
\newcommand{\emx}{\end{bmatrix}}
\newcommand{\bsmx}{\left[\begin{smallmatrix}}
\newcommand{\esmx}{\end{smallmatrix}\right]}
\newcommand{\bmxc}[1]{\left[\begin{array}{@{}#1@{}}}
\newcommand{\emxc}{\end{array}\right]}
\newcommand{\bcn}{\begin{center}}
\newcommand{\ecn}{\end{center}}

\newcommand{\Rbb}{{\mathbb{R}}}

\newcommand{\bigO}{{\mathcal{O}}}

\newcommand{\A}{\boldsymbol{A}}

\newcommand{\G}{\boldsymbol{G}}

\newcommand{\I}{\boldsymbol{I}}

\newcommand{\R}{\boldsymbol{R}}

\newcommand{\Z}{\boldsymbol{Z}}
\renewcommand{\a}{\boldsymbol{a}}

\renewcommand{\c}{\boldsymbol{c}}
\renewcommand{\d}{\boldsymbol{d}}
\newcommand{\e}{\boldsymbol{e}}
\newcommand{\f}{\boldsymbol{f}}

\newcommand{\h}{\boldsymbol{h}}

\newcommand{\p}{\boldsymbol{p}}
\newcommand{\q}{\boldsymbol{q}}

\newcommand{\s}{\boldsymbol{s}}
\newcommand{\bst}{\boldsymbol{t}} 

\renewcommand{\v}{\boldsymbol{v}}

\newcommand{\x}{{\boldsymbol{x}}}
\newcommand{\y}{{\boldsymbol{y}}}
\newcommand{\z}{\boldsymbol{z}}
\newcommand{\0}{{\boldsymbol{0}}}

\newcommand{\norm}[1]{\left\lVert #1 \right\rVert} 
\newcommand{\card}[1]{\left\lvert #1 \right\rvert} 
\newcommand{\ceil}[1]{\left\lceil #1 \right\rceil} 
\newcommand{\round}[1]{\left\lfloor #1 \right\rceil} 

\usepackage{slashbox}

\begin{document}

\title{An Efficient Algorithm for Optimally\\
Solving a Shortest Vector Problem in\\
Compute-and-Forward Design}
\author{Jinming~Wen, Baojian~Zhou,  \emph{Student Member, IEEE}, Wai Ho~Mow,  \emph{Senior Member, IEEE},  and Xiao-Wen~Chang

\thanks{This work was supported by NSERC of Canada grant 217191-12, ``Programme Avenir
Lyon Saint-Etienne de l'Universit\'e de Lyon" in the framework of the programme
``Inverstissements d'Avenir" (ANR-11-IDEX-0007), ANR through the HPAC project under Grant ANR 11 BS02 013,
and a grant from the University Grants Committee of the Hong Kong S.A.R., China (Project No. AoE/E- 02/08).
This work was presented in part at the IEEE International Conference on Communications (ICC 2015), London, UK.}
\thanks{Jinming~Wen is with the Laboratoire de l'Informatique du Parall\'elisme, (CNRS, ENS de Lyon, Inria, UCBL),
Universit\'e de Lyon, Lyon 69007, France (e-mail: jwen@math.mcgill.ca).}
\thanks{Baojian~Zhou and Wai Ho Mow are with  The Department of Electronic and Computer Engineering,
Hong Kong University of Science and Technology, Clear Water Bay, Kowloon, Hong Kong (e-mail: \{bzhouab, eewhmow\}@ust.hk).}
\thanks{X.-W. Chang is with The School of Computer Science, McGill University,
Montreal, QC H3A 2A7, Canada (e-mail: chang@cs.mcgill.ca).}}


%


\maketitle

\begin{abstract}
We consider the problem of finding the optimal coefficient vector that maximizes
the computation rate at a relay in the compute-and-forward scheme.
Based on the idea of sphere decoding, we propose a highly efficient algorithm that finds the optimal coefficient vector.
First, we derive a novel algorithm to transform the original quadratic form optimization problem into
a shortest vector problem (SVP) using the Cholesky factorization.
Instead of computing the Cholesky factor explicitly,
the proposed algorithm realizes the Cholesky factorization with only $\bigO(n)$ flops
by taking advantage of the structure of the Gram matrix in the quadratic form.
Then, we propose some conditions that can be checked with $\bigO(n)$ flops,
under which a unit vector is the optimal coefficient vector.
Finally, by taking into account some useful properties of the optimal coefficient vector,
we modify the Schnorr-Euchner search algorithm to solve the SVP.
We show that the estimated average complexity of our new algorithm is $\bigO(n^{1.5}P^{0.5})$ flops
for i.i.d. Gaussian channel entries with SNR $P$ based on the Gaussian heuristic.
Simulations show that our algorithm is not only much more efficient
than the existing ones that give the optimal solution, but also faster than some best known suboptimal methods.
Besides, we show that our algorithm can be readily adapted to output a list of $L$ best candidate vectors
for use in the compute-and-forward design.
The estimated average complexity of the resultant list-output algorithm is
$\bigO\left(n^{1.5}P^{0.5}\log L + nL\right)$ flops for i.i.d. Gaussian channel entries.
\end{abstract}


%
\IEEEpeerreviewmaketitle

\section{Introduction}

In relay networks, compute-and-forward (CF) \cite{NazG11} is a promising
relaying strategy that can offer higher rates than traditional ones (e.g.,
amplify-and-forward, decode-and-forward), especially in the moderate SNR
regime. The crucial idea of CF is the application of linear/lattice codes
\cite{Zamir2009} and physical layer network coding (PLNC) \cite{Liew2013}.
The received signal at a relay is the linear combination of a set of transmitted
signals, where the linear combination coefficients form the channel vector
from the involved sources to that relay. Through multiplying the channel vector
by an amplifying factor, the obtained new channel vector can be close to a
coefficient vector with all integer-valued entries. This means that after
applying an appropriate amplifying factor to the received signal at a relay, it
will be approximately an integer linear combination of the transmitted signals.
Since the same linear code is used at the sources, an integer linear
combination of valid codewords is still a valid codeword, which means the
aforementioned integer linear combination of the transmitted signals is
possible to be successfully decoded as a linear combination of the messages
corresponding to the transmitted signals. Under certain conditions, with
a sufficient number of such decoded linear combinations, 
the transmitted messages can be recovered.

Obviously, the amplifying factors and the integer-valued coefficient vectors need to be carefully designed.
When Nazer and Gastpar proposed the CF scheme in \cite{NazG11},
they defined the \emph{computation rate}, which
refers to the maximum transmission
rate at the involved sources of a relay such that the combined signals at the
relay can be reliably decoded.
\emph{Transmission rate}, which is the minimum computation rate over all relays,
determines the system performance.
The transmission rate becomes 0 if the
coefficient matrix formed with rows being the coefficient vectors at the relays
is not of full rank~\cite{NazG11}.
It has been pointed out that setting the amplifying factor at a relay as the
minimum-mean-square-error (MMSE) coefficient can maximize the computation rate at that relay.
The difficulty lies in the design of the coefficient vectors.
To optimize the system performance, the coefficient vectors
have to be designed jointly.
However, this requires each relay (or the destination instead) to know the
channel state information (CSI) at the other relays,
which could incur too much communication overhead in practice for large networks.
Also, the joint optimization problem could be far too complex to solve.
One alternative is to firstly develop a search algorithm to find
good coefficient vectors at each relay with the criterion being maximizing
the computation rate at that relay,
and then apply a certain strategy to coordinate relays in selection of the coefficient vectors.
This is reasonable when only the local CSI is available at each relay,
or when the network is large.
Unfortunately, the problem is difficult even for finding the coefficient vector
that maximizes the computation rate at one relay,
as it turns out to be a shortest vector problem (SVP) in a lattice.
In this paper, we shall first focus on developing the search algorithm
for finding \emph{the optimal coefficient vector at a relay}
(defined as the one that maximizes the computation rate at that relay).
After that, we will show how to adapt our algorithm
such that it can be used for solving the CF design problem.

The SVP of finding the optimal coefficient vector at a relay has attracted a lot of research interests,
and various methods have been proposed to solve the problem.
The Fincke-Pohst method \cite{FinP85} was modified in \cite{WeiC12}
to solve a different but related problem,
leading to the optimal coefficient vector and some other suboptimal vectors.
A branch-and-bound algorithm, which uses some properties of the optimal vector, was proposed in~\cite{RicSJ12}.
But it appears that this algorithm is not very efficient in this application.
There are some more efficient methods that give suboptimal solutions.
Three suboptimal methods were proposed in \cite{SakVHB12}: a method based on the complex LLL \cite{GanLM09},
a simple quantized search method, which has been improved in~\cite{SakVBH14},
and an iterative MMSE-based quantization method.
Although the average complexity of the LLL algorithm \cite{LenLL82} is polynomial if the entries of the basis vectors independently follow the
normal distribution $\mathcal{N}(0,1)$ (see, e.g., \cite{JalSM08}, \cite{LinMH13}), the complexity of the first method could be too high since
it has been proved in \cite{JalSM08} that in the MIMO context, the worst-case complexity of the LLL algorithm is not even finite.
The last two methods are of lower complexity, but they may not offer the desirable performance-complexity tradeoff, especially
when the dimension is large.
Besides these, the suboptimal quadratic programming relaxation method
in \cite{ZhoM14} and its improvement in \cite{ZhoWM14}, are of relatively low complexity.
Although their performance in terms of the computation rate are better than that of the last two methods proposed in \cite{SakVHB12},
the difference between their performance and that of the optimal methods becomes obvious as the dimension and/or the SNR get large.

In this paper (an earlier version of this paper has been posted on arXiv.org),
we propose an efficient algorithm for finding the optimal coefficient vector that maximizes the computation rate at a relay.
First, we will derive an efficient algorithm with only $\bigO(n)$ flops to transform the  problem to a SVP by
fully exploiting the structure of the Gram matrix to perform its Cholesky factorization (we do not form the whole Cholesky factor $\R$ explicitly).
Note that the complexity of the regular algorithm for Cholesky factorization is $\bigO(n^3)$.
We will also propose some conditions that can be checked with $\bigO(n)$ flops,
under which $\e_1$ (the first column of the $n\!\times\!n$ identity matrix) is the optimal coefficient vector.
Then, we will propose a modified Schnorr-Euchner search algorithm to solve the SVP
by taking advantage of the properties of the optimal solution.
Based on the Gaussian heuristic, we show that the average complexity of our new algorithm is around $\bigO(n^{1.5}P^{0.5})$ flops
for i.i.d. Gaussian channel entries with SNR $P$.
Numerical results will be given to show the effectiveness and efficiency of our algorithm.
Besides, we will show how to modify the proposed algorithm such that
it can output a list of good coefficient vectors for use in the CF design.

Preliminary results of this work have been partly presented in a conference paper \cite{WenZMC14a}.
Compared with \cite{WenZMC14a}, this work contains the following new contributions:
\begin{itemize}
\item
We use a new method to perform the Cholesky factorization to transform the optimization problem into a SVP
which reduces the complexity from $\bigO(n^3)$  to $\bigO(n)$.

\item
Some properties of the Cholesky factor $\R$ are characterized.

\item
We provide some conditions which guarantee that $\e_1$ is an optimal coefficient vector,
and these conditions can be checked with $\bigO(n)$ flops.

\item
Some new improvements on the modified Schnorr-Euchner search algorithm \cite{WenZMC14a} are made
which further accelerates the algorithm.

\item
In addition to providing more simulation results to demonstrate the effectiveness and efficiency of our algorithm,
we show that the estimated average complexity of our new algorithm is $\bigO(n^{1.5}P^{0.5})$ flops
for i.i.d. Gaussian channel entries based on the Gaussian heuristic.

\item
We show how to adapt the proposed algorithm so that it can be applied in CF design.
\end{itemize}

An algorithm with the average complexity of $\bigO(n^{2.5}P^{0.5})$ flops for i.i.d. Gaussian channel entries was proposed
in \cite{SahG14}. This algorithm finds the optimal solution by solving an optimization problem
with one variable over a bounded region, which is totally different from our proposed algorithm.
Simulations in Section \ref{sec:simcf} also indicate that our algorithm is much more efficient.

The rest of the paper is organized as follows.
In Section \ref{sec:Problem}, we start with introducing the coefficient vector design problem in CF.
Then, in Section \ref{sec:ProposedMethod}, we develop a new algorithm to solve the problem.
We analyze the complexity of our proposed method in Section \ref{sec:Complexity} and
present some numerical results in Section \ref{sec:simcf}.
In Section~\ref{sec:Adaptation}, we show how to modify our algorithm for CF design.
Finally, conclusions are given in Section \ref{sec:Conclusions}.

{\it Notation.}
Let $\mathbb{R}^n$ and $\mathbb{Z}^n$ be the spaces of the $n$-dimensional column real vectors and integer vectors, respectively.
Let $\mathbb{R}^{m\times n}$ and $\mathbb{Z}^{m\times n}$ be the spaces of the $m\times n$ real matrices and integer matrices, respectively.
Boldface lowercase letters denote column vectors and boldface uppercase letters denote matrices,
e.g., $\bst\in\mathbb{R}^n$ and $\A \in\mathbb{R}^{m\times n}$.
For a vector $\bst$, $\norm{\bst}_2$ denotes the $\ell^2$-norm of $\bst$ and $\bst^T$ denotes the transpose of $\bst$.
For $\bst\in \mathbb{R}^n$, we use $\round{\bst}$ to denote its nearest integer vector, i.e.,
each entry of $\bst$ is rounded to its nearest integer
(if there is a tie, the one with smaller magnitude is chosen).
Let $t_i$ be the element with index $i$ and $\bst_{i:j}$ be the vector composed of elements
with indices from $i$ to $j$.
$\ceil{t_i}$ denotes the smallest integer larger than or equal to $t_i$.
For a matrix $\A$, let $a_{ij}$ be the element at row  $i$ and column  $j$,
$\A_{i:j,k:\ell}$ be the submatrix containing elements with row indices from $i$ to $j$ and column indices from
$k$ to $\ell$, and
$\A_{i:j,k}$ be the vector containing elements with row indices from $i$ to $j$ and column index $k$.
Let $\0^n$ and $\0^{m\times n}$ denote the $n$-dimensional zero column vector and
$m\times n$ zero matrix, respectively.
Let $\e_k^n$ and $\textbf{1}^n$ denote the $k$-th column of an $n\!\times\!n$ identity matrix $\I$
and $n$-dimensional vector with all of it entries being 1, respectively.
Sometimes the superscripts are omitted if the dimensions are obvious.

\section{Problem Statement}
\label{sec:Problem}
We consider the problem of finding the optimal coefficient vector that
maximizes the \emph{computation rate} (defined in \cite{NazG11}) at a relay in the CF scheme.
The application scenario we focus on is the wireless relay network with slow-fading channels
and additive white Gaussian noise (AWGN).
Sources, relays, and destinations are linked with slow-fading channels,
and AWGN exists at each receiver.
For the ease of explanation, we will focus on the real-valued channel model in the sequel.

\begin{figure}[!htbp]
\centering
\includegraphics[width=180pt]{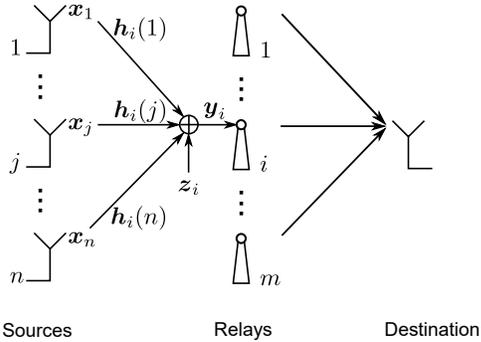}
\caption{Channel Model} \label{fig:ChannelModel}
\end{figure}

\begin{definition}
\label{def:channel}
\emph{(Channel Model)}
As shown in Figure~\ref{fig:ChannelModel},
each relay (indexed by $i=1,2,\ldots,m$) observes a noisy linear combination
of the transmitted signals through the \emph{channel},
$$\y_i = \sum_{j=1}^n \h_i(j)\x_j + \z_i,$$
where $\x_j \in \mathbb{R}^N$ with the power constraint $\frac{1}{N}\norm{\x_j}_2^2 \leq P$
is the transmitted codeword from source $j$ ($j = 1,2,\ldots,n$),
$\h_i \in \mathbb{R}^n$ is the channel vector to relay $i$
(here $\h_i(j)$ denotes the $j$-th entry of $\h_i$), $\z_i \in \mathbb{R}^N$ is the noise vector
with entries being i.i.d. Gaussian, i.e., $\z_i \sim \mathcal{N}\left(\0,\I\right)$,
and $\y_i$ is the signal received at relay $i$.
\end{definition}

For relay $i$ with the channel vector $\h_i$, let $\a_i$ be the chosen
coefficient vector, the computation rate is calculated according to the following
theorem~\cite{NazG11}.

\begin{theorem}
\label{theorem:ComputationRate}
The computation rate at relay $i$ is uniquely maximized by choosing the amplifying
factor as the MMSE coefficient, which results in a computation rate
\begin{equation}
\begin{aligned}
\label{e:ComputationRate}
\mathcal{R} \left( \h_i, \a_i \right) =
\frac{1}{2} \log^+
\left( \frac{1}{
\norm{\a_i}^2 -
\frac{P\left( \h_i^T\a_i \right)^2}{1 + P\norm{\h_i}_2^2}
} \right),
\end{aligned}
\end{equation}
where the $\log$ function is with respect to base 2 and $\log^+(x) \triangleq \max \left( \log(x),0 \right)$.
\end{theorem}

Also, we define the optimal coefficient vector for a relay as below.

\begin{definition}
\emph{(The Optimal Coefficient Vector)}
The \emph{optimal coefficient vector} $\a_i^\star$ for a relay with
channel vector $\h_i$ is the one that maximizes the computation rate,
\begin{equation}
\begin{aligned}
\label{e:aOpt}
\a_i^\star = \arg \max_{\a_i \in \mathbb{Z}^n \backslash\{\0\}} \mathcal{R} \left( \h_i, \a_i \right).
\end{aligned}
\end{equation}
\end{definition}

The optimization problem \eqref{e:aOpt} can be further formulated as the following problem \cite{WeiC12}:
\begin{subequations}\label{e:CFPmodel}
\begin{align}
\a_i^\star &=\arg\min_{\a_i\in\mathbb{Z}^n\backslash\{\0\}} \a_i^T\G_i\a_i\label{e:CFP}, \\
 \G_i &=\I - \frac{P}{1 + P\norm{\h_i}_2^2} \h_i\h_i^T.
\label{e:Gi}
\end{align}
\end{subequations}

Hereafter, we will focus on relay $i$, and thus ignore the subscript ``$i$'', e.g., $\h_i$ will be directly written as $\h$.
In the next section, an efficient method to solve \eqref{e:CFPmodel} based on sphere decoding will be provided.

\section{Proposed Method}
\label{sec:ProposedMethod}

Define the \emph{scaled channel vector} $\bst$ as
\begin{equation}
\label{e:t}
\bst=\sqrt{\frac{P}{1 + P\norm{\h}_2^2}}\h.
\end{equation}
Then, \eqref{e:CFPmodel} is equivalent to the following problem:
\begin{subequations}\label{e:CFmodel}
\begin{align}
\a^\star &= \arg\min_{\a\in\mathbb{Z}^n\backslash\{\0\}}\a^T\G\a, \label{e:CFPT} \\
\G &\triangleq \I - \bst \bst^T.\label{e:G}
\end{align}
\end{subequations}
Obviously, $\norm{\bst}_2 < 1$ and $\G$ is symmetric positive definite.
Throughout this paper, we assume $\h\neq\0$; otherwise, it is trivial.

The problem in \eqref{e:CFmodel} can be solved via the following two steps:
\begin{itemize}
\item First, for a given $\bst$, compute $\G$ and find its Cholesky factorization, i.e., find an upper triangular matrix $\R$ such that  $\G=\R^T\R$.
Then \eqref{e:CFmodel} is equivalent to the following shortest vector problem (SVP),
\begin{equation}
\label{e:SVP}
\a^\star = \arg\min_{\a\in\mathbb{Z}^n\backslash\{\0\}}\norm{\R\a}_2.
\end{equation}

\item Second, use a search algorithm, such as the LLL-aided  Schnorr-Euchner
search strategy \cite{SchE94}, to solve \eqref{e:SVP}. We will explain the details later.
\end{itemize}
It is easy to see that for a given $\bst$, computing $\G$ costs $\bigO (n^2)$ flops.
Besides, it is well-known that computing the Cholesky factorization of a general $n\times n$ matrix costs $\bigO (n^3)$ flops.
Moreover, the complexity of the LLL-aided  Schnorr-Euchner search strategy \cite{SchE94} for solving \eqref{e:SVP} may be too high.
Fortunately, we find out that it is possible to accelerate the aforementioned two steps as follows:
\begin{itemize}
\item First, take advantage of the special structure of $\G$ in \eqref{e:G}  to compute its Cholesky factorization
and transform \eqref{e:CFmodel} to \eqref{e:SVP}, but do not explicitly form $\G$, the whole $\R$ and the SVP.

\item Second, investigate the properties of a solution $\a^{\ast}$ to \eqref{e:SVP} and take them
into account to modify the Schnorr-Euchner search strategy \cite{SchE94} to find $\a^\star$.
\end{itemize}

Obviously, if $\a^\star$ is a solution of \eqref{e:SVP}, so is $-\a^\star$.
For simplicity, we apply the following restriction.
\begin{Restriction}
\label{res:optimal}
Throughout this paper, we restrict the solution $\a^\star$ to \eqref{e:SVP} such that $\bst^T\a^\star\geq0$.
\end{Restriction}

\subsection{Transformation of the problem}
\label{sec:formR}

To transform \eqref{e:CFmodel} to the SVP \eqref{e:SVP}, we need to find the  Cholesky factor $\R$ of $\G$ in \eqref{e:G}.
Besides the regular method, one can use the algorithm proposed in \cite{BojBVD87},
which costs $2n^2+\bigO(n)$ flops, to get the Cholesky factor $\R$. However, the complexity can be further reduced.
Also, to analyze the complexity of our proposed search algorithm in Section \ref{sec:Complexity},
we need to know the diagonal entries of $\R$.
In this subsection, we will take into account the special structure of $\G$ to achieve this goal
with only $\bigO(n)$ flops (we do not form the whole $\R$ explicitly. If the whole $\R$ is needed
for other applications, it costs $n^2/2+\bigO(n)$ flops).
Based on the diagonal entries of $\R$ and by investigating their properties,
we will also propose some conditions that can be checked with $\bigO(n)$ flops,
under which the optimal solution $\a^\star$ can be obtained immediately without using any search algorithm.

Our proposed algorithm to find the Cholesky factor $\R$ of $\G$ in \eqref{e:G} is based on the following theorem:

\begin{theorem}
\label{t:chol}
The Cholesky factor $\R$ of $\G$ in \eqref{e:G} is given by:
\begin{eqnarray}
\label{e:Ri}
r_{ij}=
\begin{cases}
\sqrt{\frac{1-\sum_{l=1}^{i}t^2_l}{1-\sum_{l=1}^{i-1}t^2_l}}, & j=i \cr
\frac{-t_it_j}{\sqrt{1-\sum_{l=1}^{i-1}t^2_l}\sqrt{1-\sum_{l=1}^{i}t^2_l}}, & i<j\leq n \cr
 \end{cases}
,
\end{eqnarray}
where $1\leq i\leq n$ and denote $\sum_{1}^0\cdot=0$.
\end{theorem}
{\bf Proof.}
To prove the theorem, we  show any element of $\G$ is equal to the corresponding element of
$\R^T\R$ in the same position, i.e., by \eqref{e:G},  we would like to show
\beq
\label{e:diag}
\sum_{i=1}^{k}r_{ik}^2=1-t_{k}^2, \;  \, 1\leq k\leq n
\eeq
and
\beq
\label{e:offdiag}
\sum_{i=1}^{k}r_{ik}r_{ij}=-t_{k}t_j,\;  \, 1\leq k<j\leq n.
\eeq

By \eqref{e:Ri}, we have
\begin{align*}
&\sum_{i=1}^{k}r_{ik}^2\\
=\;&r_{kk}^2+\sum_{i=1}^{k-1}r_{ik}^2\\
=\;&\frac{1-\sum_{l=1}^{k}t^2_l}{1-\sum_{l=1}^{k-1}t^2_l}
+\sum_{i=1}^{k-1}\frac{t_i^2t_{k}^2}{(1-\sum_{l=1}^{i-1}t^2_l)(1-\sum_{l=1}^{i}t^2_l)}\\
=\;&\frac{1-\sum_{l=1}^{k}t^2_l}{1-\sum_{l=1}^{k-1}t^2_l}
+t_{k}^2\sum_{i=1}^{k-1}\big(\frac{1}{1-\sum_{l=1}^{i}t^2_l}-\frac{1}{1-\sum_{l=1}^{i-1}t^2_l}\big)\\
=\;&\frac{1-\sum_{l=1}^{k}t^2_l}{1-\sum_{l=1}^{k-1}t^2_l}
+t_{k}^2\big(\frac{1}{1-\sum_{l=1}^{k-1}t^2_l}-1\big)\\
=\;&\frac{1-\sum_{l=1}^{k}t^2_l}{1-\sum_{l=1}^{k-1}t^2_l}
+\frac{t_{k}^2\sum_{l=1}^{k-1}t_l^2}{1-\sum_{l=1}^{k-1}t_l^2}=1-t_k^2,
\end{align*}
and
\begin{align*}
&\sum_{i=1}^{k}r_{ik}r_{ij}\\
=\;&r_{kk}r_{kj}+\sum_{i=1}^{k-1}r_{ik}r_{ij}\\
=\;&\frac{-t_{k}t_j}{1-\sum_{i=1}^{k-1}t_i^2}+
\sum_{i=1}^{k-1}\frac{t_i^2t_{k}t_j}{(1-\sum_{l=1}^{i-1}t^2_l)(1-\sum_{l=1}^{i}t^2_l)}\\
=\;&\frac{-t_{k}t_j}{1-\sum_{i=1}^{k-1}t_i^2}+
t_{k}t_j\sum_{i=1}^{k-1}\frac{t_i^2}{(1-\sum_{l=1}^{i-1}t^2_l)(1-\sum_{l=1}^{i}t^2_l)}\\
=\;&\frac{-t_{k}t_j}{1-\sum_{i=1}^{k-1}t_i^2}+
t_{k}t_j\big(\frac{1}{1-\sum_{l=1}^{k-1}t^2_l}-1\big)\\
=\;&\frac{-t_{k}t_j}{1-\sum_{i=1}^{k-1}t_i^2}+
\frac{t_{k}t_j\sum_{i=1}^{k-1}t_i^2}{1-\sum_{i=1}^{k-1}t_i^2}=-t_kt_j.
\end{align*}
Thus, both \eqref{e:diag} and \eqref{e:offdiag} hold, completing the proof.
 \ \ $\Box$

We can use Theorem \ref{t:chol} to design an efficient algorithm to find $\R$. To simplify notation,
we introduce an $(n+1)$-dimensional vector variable $\f$. Let
\beq
\label{e:fi}
f_0=1, \quad f_i=1-\sum_{l=1}^{i}t^2_l, \quad 1\leq i\leq n.
\eeq
Then by \eqref{e:Ri}, we have
$$
r_{ii}=\sqrt{f_i/f_{i-1}}, \quad 1\leq i\leq n,
$$
and
$$
\R_{i,i+1:n}=(-t_i/\sqrt{f_if_{i-1}})\bst^T_{i+1:n}, \quad 1\leq i< n.
$$
Note that $\R_{i, i + 1 : n}$ is a scaled $\bst^T_{i+1:n}$.

After getting $\R$, we will modify the Schnorr-Euchner search algorithm to solve \eqref{e:SVP}.
Later we will see that it is not necessary to form $\R$ explicitly
(we will give more details to explain this in Section \ref{ss:Msearch}), i.e, we
do not need to compute the multiplication of $-t_i/\sqrt{f_if_{i-1}}$ and $\bst_{i+1:n}$.
Thus, the complexity of obtaining $\R$ is $\bigO(n)$ flops.

By \eqref{e:Ri}, it is easy to see that $\R$ has the following properties
which are useful to analyze the complexity of our proposed search algorithm.
\begin{theorem}
\label{t:propertyr}
For $1\leq k \leq n$, the following inequalities hold:
\beq
\label{ine:propertyr2}
\sqrt{1-\sum_{i=1}^kt_i^2}\leq r_{kk}\leq\sqrt{1-t_k^2},
\eeq
\beq
\label{e:Dk}
\prod_{i=k}^nr_{ii}=\frac{\sqrt{1-\norm{\bst}_2^2}}{\sqrt{1-\sum_{i=1}^{k-1}t_i^2}}
\geq\sqrt{1-\norm{\bst}_2^2}.
\eeq
\end{theorem}
{\bf Proof.}
The first inequality in \eqref{ine:propertyr2} follows direct from \eqref{e:Ri} and
the basic fact that $1-\sum_{l=1}^{i-1}t_l^2\leq 1$ for $1\leq i\leq n$ (recall that we define $\sum_{l=1}^{0}\cdot =0$).

The second inequality in \eqref{ine:propertyr2} follows direct from \eqref{e:Ri} and some basic calculations.

The equality in \eqref{e:Dk} follows direct from \eqref{e:Ri}, and the inequality in \eqref{e:Dk} follows  from
the basic fact that $1-\sum_{l=1}^{i-1}t_l^2\leq 1$ for $1\leq i\leq n$.
 \ \ $\Box$

By Theorem \ref{t:chol}, we have the following interesting result, which can be used to describe the geometry of the search space later.
\begin{theorem}
\label{t:eig}
For $1\leq i<j\leq n$, the eigenvalues of $\R_{i:j,i:j}^T\R_{i:j,i:j}$ are 1 with algebraic multiplicity $j-i$
and $f_j/f_{i-1}$, where $\f$ is defined in \eqref{e:fi}.
\end{theorem}
{\bf Proof.} We first prove
\beq
\label{e:Gin}
\R_{i:n,i:n}^T\R_{i:n,i:n}=\I_{n-i+1}-\frac{\bst_{i:n}}{\sqrt{f_{i-1}}}\frac{\bst_{i:n}^T}{\sqrt{f_{i-1}}}.
\eeq
If $i=1$, then by Theorem \ref{t:chol} and \eqref{e:fi}, \eqref{e:Gin} holds. So we only need to prove it holds for $i>1$.

By Theorem \ref{t:chol}, we have
\begin{align*}
&\bmx
\R^T_{1:i-1,1:i-1}&\0\\
\R_{1:i-1,i:n}^T&\R_{i:n,i:n}^T
\emx
\bmx
\R_{1:i-1,1:i-1}&\R_{1:i-1,i:n}\\
\0&\R_{i:n,i:n}
\emx\\
=&
\bmx
\I_{i-1}&\0\\
\0&\I_{n-i+1}
\emx
-
\bmx
\bst_{1:i-1}\bst_{1:i-1}^T&\bst_{1:i-1}\bst_{i:n}^T\\
\bst_{i:n}\bst_{1:i-1}^T&\bst_{i:n}\bst_{i:n}^T
\emx.
\end{align*}
The right bottom parts of both sides are the same, thus,
$$
\R_{i:n,i:n}^T\R_{i:n,i:n}=\I_{n-i+1}-\bst_{i:n}\bst_{i:n}^T-\R_{1:i-1,i:n}^T\R_{1:i-1,i:n}.
$$
By Theorem \ref{t:chol} and \eqref{e:fi}, we have
\begin{align*}
\R_{1:i-1,i:n}=
(\bst_{i:n}
\bmx
\frac{-t_1}{\sqrt{f_0f_1}}& \frac{-t_2}{\sqrt{f_1f_2}}&\ldots&\frac{-t_{i-1}}{\sqrt{f_{i-2}f_{i-1}}}
\emx)^T.
\end{align*}
Thus, we obtain
$$
\R_{i:n,i:n}^T\R_{i:n,i:n}=\I_{n-i+1}-(1+\sum_{k=1}^{i-1}\frac{t_k^2}{f_kf_{k-1}})\bst_{i:n}\bst_{i:n}^T.
$$
By \eqref{e:fi},
$$
\frac{t_k^2}{f_kf_{k-1}}=\frac{1}{f_{k}}-\frac{1}{f_{k-1}}.
$$
Therefore,
$$
\R_{i:n,i:n}^T\R_{i:n,i:n}=\I_{n-i+1}-\frac{\bst_{i:n}}{\sqrt{f_{i-1}}}\frac{\bst_{i:n}^T}{\sqrt{f_{i-1}}},
$$
i.e., \eqref{e:Gin} holds.

From \eqref{e:Gin}, we can immediately get
$$
\R_{i:j,i:j}^T\R_{i:j,i:j}=\I_{j-i+1}-\frac{\bst_{i:j}}{\sqrt{f_{i-1}}}\frac{\bst_{i:j}^T}{\sqrt{f_{i-1}}}.
$$
Thus, the eigenvalues of $\R_{i:j,i:j}^T\R_{i:j,i:j}$ are 1's and
$$
1-\frac{\sum_{k=i}^jt_k^2}{f_{i-1}}=\frac{f_{j}}{f_{i-1}}.
$$
\ \ $\Box$

Generally speaking, after getting $\R$, a search algorithm should be used to find the solution $\a^\star$ to \eqref{e:SVP}.
Theorem \ref{t:chol} gives the closed-form expression of $\R$, so a natural question is
whether there exist some easily-checked conditions,
under which the optimal solution $\a^\star$ can be obtained without using any search algorithm?
In the following, we will answer this question.
\begin{theorem}
\label{t:LBDs}
The optimal solution $\a^\star$ satisfies
\begin{eqnarray}
\label{ine:astar}
\|\R\a^\star\|_2 \geq \min_{1\leq i\leq n} \sqrt{\frac{1-\sum_{j=1}^{i}t_j^2}{1-\sum_{j=1}^{i-1}t_j^2}} \geq \sqrt{1-\norm{\bst}_2^2}.
\end{eqnarray}
Furthermore, if we have
\begin{eqnarray}
\label{e:closdexp}
t_i^2\leq t_1^2(1-\sum_{j=1}^{i-1}t_j^2), \quad \,\;i=2, 3\ldots,n,
\end{eqnarray}
then $\e_1$ is a solution to \eqref{e:SVP}.
\end{theorem}
{\bf Proof.}
The first inequality  in \eqref{ine:astar} follows directly from \eqref{e:Ri} and
\begin{eqnarray}
\label{ine:astar2}
\|\R\a^\star\|_2 \geq \min_{1\leq i\leq n} r_{ii},
\end{eqnarray}
which was given in \cite[pp.99]{DanG02}.

By the first inequality  in \eqref{ine:propertyr2},
$$
\min_{1\leq i\leq n}r_{ii}\geq \sqrt{1-\norm{\bst}_2^2}.
$$
Therefore, the second inequality in \eqref{ine:astar} follows.

In the following, we prove the second part.
If for some $1\leq i\leq n$,
\[
t_i^2\leq t_1^2(1-\sum_{j=1}^{i-1}t_j^2).
\]
Then,
\begin{align*}
\sum_{j=1}^{i}t_j^2-\sum_{j=1}^{i-1}t_j^2\leq t_1^2(1-\sum_{j=1}^{i-1}t_j^2).
\end{align*}
Thus,
\begin{align*}
1-\sum_{j=1}^{i}t_j^2\geq (1-\sum_{j=1}^{i-1}t_j^2)- t_1^2(1-\sum_{j=1}^{i-1}t_j^2).
\end{align*}
Equivalently, we have
\[
\sqrt{\frac{1-\sum_{j=1}^{i}t_j^2}{1-\sum_{j=1}^{i-1}t_j^2}}\geq \sqrt{1-t_1^2}.
\]
Thus, from \eqref{e:closdexp} holds, we have
\begin{eqnarray*}
\min_{1\leq i\leq n}\sqrt{\frac{1-\sum_{j=1}^{i}t_j^2}{1-\sum_{j=1}^{i-1}t_j^2}}=\sqrt{1-t_1^2}.
\end{eqnarray*}
Therefore, the first inequality in \eqref{ine:astar} becomes an equality with $\a^\star=\e_1$,
so $\e_1$ is a solution to \eqref{e:SVP}.
\ \ $\Box$

It is easy to see that \eqref{e:closdexp} can be checked by $\bigO(n)$ flops.

\begin{remark}
From \eqref{e:G}, we can see that
\[
\min_{1\leq i\leq n}\e_i^T\G\e_i=\min_{1\leq i\leq n}(\|\e_i\|^2_2-(\e_i^T\bst)^2)=\min_{1\leq i\leq n}(1-t_i^2).
\]
Thus, if $\max\limits_{1\leq i\leq n} |t_i|\neq |t_1|$, then $\e_1$ cannot be an optimal solution.
Thus, \eqref{e:closdexp} does not hold.
However, we can order the entries of $\bst$ such that after the ordering, $\max\limits_{1\leq i\leq n} |t_i|=|t_1|$.
Clearly, doing this can increase the probability of \eqref{e:closdexp} holds.
And $\e_j$ with $j$ satisfying $\max_{1\leq i\leq n} |t_i|= |t_j|$ ( here $\bst$ is the vector before using the transformation) is an optimal solution.
Naturally, it is interesting to know how often does \eqref{e:closdexp} hold?
We will do some simulations for this in the end of next subsection.
\end{remark}

\subsection{Reordering the entries of $\bst$}
\label{a:Reordering}
After getting \eqref{e:SVP}, a search algorithm, such as the Schnorr-Euchner
search strategy \cite{SchE94} can be used to solve it, i.e., finding the shortest nonzero vector of the lattice
$\mathcal{L}(\R)$, which is defined by
\[
\mathcal{L}(\R)=\{\R\z|\z \in \mathbb{Z}^n\}.
\]
The columns of $\R$  form a basis of $\mathcal{L}(\R)$
(note that the basis of a lattice is not necessary an upper triangular matrix, but it must be a full column rank matrix).
For any $n\geq2$, $\mathcal{L}(\R)$ has infinity many  bases and any of two are connected by a unimodular matrix $\Z$,
i.e.,  $\Z \in \mathbb{Z}^{n\times n}$ and $\det(\Z)=\pm1$.  Specifically, for each given lattice basis matrix
$\R\in \mathbb{R}^{m\times n}$, $\R\Z$ is also a basis matrix of $\mathcal{L}(\R)$ if and only if $\Z$ is unimodular,
see, e.g., \cite{AgrEVZ02}.
The process of selecting a good basis for a given lattice, given some criterion, is called lattice reduction.
In many applications, it is advantageous if the basis vectors are short
and close to be orthogonal \cite{AgrEVZ02}.
For more than a century, lattice reduction have been investigated by many people and several  types of reductions
have been proposed,  including the KZ reduction \cite{KZ73} (an efficient KZ reduction algorithm can be found in \cite{WenC15}),
the Minkowski reduction \cite{Min96} (an efficient Minkowski reduction algorithm can be found in \cite{ZhaQW12}),
the LLL reduction \cite{LenLL82} and Seysen's reduction \cite{Sey93} etc.

For efficiency, lattice reduction for $\R$ in \eqref{e:SVP}
is usually used to strive for
\begin{eqnarray}
\label{ine:orderR}
r_{11}\leq r_{22}\leq\ldots \leq r_{nn}
\end{eqnarray}
to accelerate searching. Notice that \eqref{ine:orderR} may not be achievable. For more details on why \eqref{ine:orderR} is desirable for,
readers are referred to, e.g., \cite{AgrEVZ02} and \cite{ChaWX13}.

The LLL reduction \cite{LenLL82} is a commonly used reduction method to strive for \eqref{ine:orderR}. However, for this application,
it has two main drawbacks. First, its complexity is high. In fact, it was shown in \cite{JalSM08} that in the MIMO context,
the worst-case complexity is not even finite. For more details, see, e.g., \cite{LenLL82}, \cite{DauV94} and \cite{LinMH13}.
Also, from the simulation results in Section \ref{sec:simcf}, we will see that
the complexity of the LLL reduction is even higher than that of our proposed algorithm.
Second, it may destroy the structure of $\R$ and some properties of the optimal solution $\a^\star$
to the reduced problem (we will explain this in the latter part of this subsection).
In this subsection, we will propose a  method to strive for \eqref{ine:orderR} without the above shortcomings.

From \eqref{e:Ri}, to strive for \eqref{ine:orderR}, we permute the entries of $\bst$.
To make $r_{11}$ as small as possible, we permute $\bst$ such that $|t_1|$ is the  largest.
Suppose that $t_j, 1\leq j\leq i$ have been fixed,
then from \eqref{e:Ri}, $r_{jj}, 1\leq j\leq i$ are fixed. To make $r_{j+1,j+1}$ as small as possible,
we permute the entries of $t_j, i+1\leq j\leq n$ such that $|t_{j+1}|$ is the largest.
So after the permutations we have
\begin{align}
\label{e:abstOrdered}
|t_1|\geq |t_2|\geq \ldots \geq |t_n|.
\end{align}

Here we want to point out the above idea  of reordering the entries of $\bst$ is actually the same as that of SQRD \cite{WubBRKK01},
a column reordering strategy for a general matrix in the box-constrained integer least squares (BILS) problem \cite{ChaH08}, \cite{WenC14}.
It is interesting to note that if we use the idea of V-BLAST \cite{FosGVW99}, another column reordering strategy used in
solving BILS problems \cite{DamGC03}, we will get the same ordering of $\bst$.  In fact, by \eqref{e:Ri},
$$
r_{nn}^2=\frac{1-\norm{\bst}_2^2}{1-\norm{\bst}_2^2+t_n^2}.
$$
Thus, to make $r_{nn}$ as large as possible, we need to permute $\bst$ such that $|t_n|$ is the smallest.
Suppose that $t_j, i+1\leq j\leq n$ have been fixed, then from \eqref{e:Ri}, $r_{jj},i+1\leq j\leq n$ are fixed.
By \eqref{e:Ri},
$$
r_{ii}^2=\frac{1-\norm{\bst}_2^2+\sum_{j=i+1}^nt_j^2}{1-\norm{\bst}_2^2+\sum_{j=i+1}^nt_j^2+t_i^2}.
$$
Thus, to make $r_{jj}$ as large as possible, we permute the entries of $t_j, 1\leq j\leq i$ such that $|t_{i}|$ is the smallest.
So after the permutations we also have \eqref{e:abstOrdered}.

To make the search process faster, we also want to make $t_i\geq0$ for $1\leq i\leq n$.
This can easily be done. In fact, when we determine the $i$-th entry of $\bst$ in the permutation process,
we can use a sign permutation matrix so that the new $i$-th entry is nonnegative. Thus, eventually we have
\begin{align}
\label{e:tOrdered}
t_1\geq t_2\geq \ldots \geq t_n\geq 0.
\end{align}

The above process can be described mathematically as follows.
For any given $\bst$, it is easy to find a signed permutation  matrix $\Z\in \mathbb{Z}^{n\times n}$ such that $\bar{\bst}=\Z \bst$ satisfying:
\begin{equation*}
\bar{t}_1\geq \bar{t}_2\geq\ldots \geq \bar{t}_n\geq0.
\end{equation*}
This transformation is a sorting process and the complexity is $\bigO(n\log(n))$, see \cite{ZhoM14} for more details.
Note that $\Z\Z^T=\I$. Then, with $\bar{\a}=\Z\a$, the optimization problem \eqref{e:CFmodel} can be transformed to
\begin{align*}
\bar{\a}^\star &= \min_{\bar{\a}\in\mathbb{Z}^n\backslash\{\0\}}\bar{\a}^T\bar{\G}\bar{\a},\\
\bar{\G} &\triangleq \I - \bar{\bst} \bar{\bst}^T.
\end{align*}
Obviously $\a^\star=\Z^{T}\bar{\a}^\star$.

Therefore, for the sake of convenience, in our later analysis, we assume $\bst$ satisfies \eqref{e:tOrdered}.

In addition to speeding up the search, ordering the entries of $\bst$ like in \eqref{e:tOrdered} has another important effect, i.e.,
by the results in \cite{RicSJ12} and \cite{ZhoM14}, if \eqref{e:tOrdered} holds, we can find a solution
$\a^\star$ to \eqref{e:SVP} such that
\begin{align}
\label{e:aOrdered}
a^\star_1\geq a^\star_2\geq \ldots\geq a^\star_n \geq 0.
\end{align}

The order of the elements of the solution  $\a^\ast$ in \eqref{e:aOrdered} is a key property of the solution we restricted for \eqref{e:SVP}.
It has been used in \cite{ZhoM14} to find a suboptimal solution to \eqref{e:SVP}, but only the property that $a_i\geq0, 1\leq i\leq n$
has been used in \cite{RicSJ12} to solve \eqref{e:SVP}. In this paper, we will take full advantage of it in designing the search algorithm.
Note that, if the LLL reduction is used for reducing $\R$ in \eqref{e:SVP}, then \eqref{e:aOrdered} may not hold, which is the second drawback of using
the LLL reduction  in striving for \eqref{e:tOrdered}.
The motivation for reordering the entries of $\bst$ in  \cite{RicSJ12} and \cite{ZhoM14} is to obtain the property \eqref{e:aOrdered},
which was (partially) used in their methods. Here we gave another motivation from the search point of view.

Under \eqref{e:tOrdered} and Theorem \ref{t:chol}, we have the following interesting results:
\begin{theorem}
\label{t:orderRnorm}
If \eqref{e:tOrdered} holds, then for $1\leq i\leq n-1$
\begin{align}
\label{e:orderRnorm}
r_{ii}\leq\norm{\R_{i:i+1,i+1}}_2\leq\norm{\R_{i:i+2,i+2}}_2\leq \ldots\leq \norm{\R_{i:n,n}}_2
\end{align}
\end{theorem}
{\bf Proof.} By \eqref{e:Ri}, for $1\leq i<j\leq n$,
\begin{align*}
&\norm{\R_{i:j,j}}^2\\
=&\sum_{k=i}^{j-1}r_{kj}^2+r_{jj}^2\\
=&\sum_{k=i}^{j-1}\frac{t_k^2t_j^2}{(1-\sum_{l=1}^{k-1}t_l^2)(1-\sum_{l=1}^{k}t_l^2)}+
\frac{(1-\sum_{l=1}^{j}t_l^2)}{(1-\sum_{l=1}^{j-1}t_l^2)}\\
=&\frac{t_j^2}{(1-\sum_{l=1}^{j-1}t_l^2)}-\frac{t_j^2}{(1-\sum_{l=1}^{i-1}t_l^2)}+
\frac{(1-\sum_{l=1}^{j}t_l^2)}{(1-\sum_{l=1}^{j-1}t_l^2)}\\
=&1-\frac{t_j^2}{(1-\sum_{l=1}^{i-1}t_l^2)}.
\end{align*}

By the aforementioned equations, \eqref{e:Ri} and \eqref{e:tOrdered}, it is easy to see that \eqref{e:orderRnorm} holds.
$\Box$

From Theorem \ref{t:LBDs} we can see that if \eqref{e:closdexp} holds, $\e_1$ is a solution to \eqref{e:SVP}.
In the following, we do some simulations to show how often does \eqref{e:closdexp} hold?
For each $n$ and $P$, we randomly generate $10000$ realizations of $\h$.
Then, we compute $\bst$ by \eqref{e:t} and transform it such that \eqref{e:tOrdered} holds.
Tables \ref{tb:optimal1} and \ref{tb:optimal2} respectively show the total number of
cases over 10000 realizations that \eqref{e:closdexp} holds for
$n$ from 2 to 16 with step 2 and $n$ from 100 to 800 with step 100.
From Tables \ref{tb:optimal1} and \ref{tb:optimal2}, we can see that,
\eqref{e:closdexp} holds with a high probability when both $n$ and $P$ are small;
however, it holds with a very low probability when both $n$ and $P$ are very large.

\begin{table}[tbp!]
\caption{Number of $e_1$ being the optimal solution over $10000$ realizations of $\h$}
\scriptsize
\centering
\begin{tabular}{|c|c|c|c|c|c|c|c|c|c||}
 \hline
\backslashbox{$P$}{$n$}& 2&4&6&8&10&12&14&16\\ \hline
P=0 dB  & 8617 &       6172 &       4948   &     4255   &     3778  &      3641   &     3486 &       3468 \\ \hline
P=10 dB  & 6148  &      2767 &       1728  &      1222   &     1025  &       881 &        790 &        731\\ \hline
P=20 dB  &  3903      &   944     &    413    &     223       &  146    &     101     &     70     &     59\\ \hline
\end{tabular}
\label{tb:optimal1}
\end{table}

\begin{table}[tbp!]
\caption{Number of $e_1$ being the optimal solution over $10000$ realizations of $\h$}
\scriptsize
\centering
\begin{tabular}{|c|c|c|c|c|c|c|c|c|c||}
 \hline
\backslashbox{$P$}{$n$}& 100&200&300&400&500&600&700&800\\ \hline
P=0 dB  & 4837 &       4833   &     4568 &       4065  &      3540  &      3324  &      2866  &      2589 \\ \hline
P=10 dB  & 196   &       64  &        35  &        15 &          9  &         5   &        9    &       2\\ \hline
P=20 dB  &  0      &   0     &    0   &     0      &  0    &     0     &     0     &     0\\ \hline
\end{tabular}
\label{tb:optimal2}
\end{table}

\subsection{Schnorr-Euchner search algorithm}

We first introduce a depth-first tree search algorithm:
the Schnorr-Euchner search algorithm \cite{SchE94}, \cite{AgrEVZ02}, a variation of the Fincke-Pohst search strategy \cite{FinP85},
to solve  a general SVP,  which has the form of \eqref{e:SVP}.
Note that, the Schnorr-Euchner algorithm is generally more efficient than the Fincke-Pohst,
for more details, see, e.g., \cite{AgrEVZ02}.
Then we modify it by using the properties of $\R$ and the optimal solution $\a^\star$ to make the search process faster.

Let the optimal solution be within the following hyper-ellipsoid:
\begin{eqnarray}
\label{ine:ellp}
\norm{\R\a}_2^2< \beta^2,
\end{eqnarray}
where $\beta$ is a constant. Define
\begin{eqnarray}
\label{e:d}
d_n=0, \; d_k=-\frac{1}{r_{kk}}\sum_{j=k+1}^nr_{kj}a_j, \quad k=n-1,\ldots, 1.
\end{eqnarray}
Then \eqref{ine:ellp} can be written as:
$$
\sum_{i=1}^nr_{ii}^2(a_i-d_i)^2< \beta^2
$$
which is equivalent to
\begin{eqnarray}
\label{e:levelk}
r_{kk}^2(a_k-d_k)^2< \beta^2-\sum_{j=k+1}^nr_{jj}^2(a_i-d_j)^2
\end{eqnarray}
for $k=n, n-1,\ldots, 1$, where $k$ is called the level index and $\sum_{j=n+1}^n\cdot=0$.

Based on \eqref{e:levelk}, the Schnorr-Euchner search algorithm can be described as follows.
First we set the initial $\beta=\infty$, and for $k=n, n-1,\ldots, 1$, we compute $d_{k}$ by \eqref{e:d} and set
$a_k=\lfloor d_k\rceil$, leading to $a_k=0$, for which  \eqref{e:levelk} holds. So we obtain an integer vector $\a=\0$.
Since the optimal solution $\a^\star$ is a nonzero vector, we need to update $\a$. Specifically, we set $a_1$ as the next closest integer to $d_1$.
Note that \eqref{e:levelk} with $k=1$ holds for the updated $\a$.
Then, we store this updated $\a$ and set $\beta=\norm{\R\a}_2$.
After this, we try to find an integer vector within the new ellipsoid by updating the latest found $\a$.
Obviously, we cannot update only its first entry $a_1$, since we cannot
find any new integer $a_1$ that satisfies \eqref{e:levelk} with $k=1$, which is now an equality for the current $\a$.
Thus we move up to level 2 to try to update $a_2$ by choosing it being the next nearest integer to $d_2$.
If it satisfies \eqref{e:levelk} with $k=2$, we move down to level 1 to  update $a_1$
by computing $d_1$ (see \eqref{e:d}) and setting $a_1 = \lfloor d_1 \rceil$ and then checking if \eqref{e:levelk} with $k=1$ holds and so on;
otherwise we move up to level 3 to try to update $a_3$, and so on. Finally, when we fail to find a new value for
$a_n$ to satisfy \eqref{e:levelk} with $k=n$, the search process stops and the latest
found integer vector is the optimal solution $\a^\star$ we seek.
This is a depth-first tree search. For more details, see, e.g., \cite{AgrEVZ02} and \cite{ChaH08}.

We summarize the search process in Algorithm \ref{a:search}, where
\begin{eqnarray}
\label{e:sgn}
\mbox{sgn}(x)=
\begin{cases}
1, & x\geq0 \cr
-1, &x<0
 \end{cases}.
\end{eqnarray}

\begin{algorithm}[!ht]
\caption{ Schnorr-Euchner search algorithm}
\label{a:search}
\textbf{Input:}\ \ \ A nonsingular upper triangular matrix $\R\in \mathbb{R}^{n\times n}$\\
\textbf{Output:} A solution $\a^\star$ to the SVP in \eqref{e:SVP}

\begin{enumerate}
\item (Initialization) Set $k=n,\beta=+\infty$.
\item Compute $d_k$ by using \eqref{e:d}, set $a_k=\lfloor d_k\rceil$ and $s_k=\mbox{sgn}(d_k-a_k)$ (see \eqref{e:sgn}).
\item (Main Step) If the inequality in \eqref{e:levelk} does not hold, then go to Step 4. Else if $k>1$,
 set $k=k-1$ and go to Step 2.
Else ($k=1$), go to Step 5.
\item (Outside ellipsoid) If $k=n$, terminate. Else, set $k=k+1$ and go to Step 6.
\item \label{i:level1} (A valid point is found) If $\a$ is a nonzero vector, then save $\a^\star=\a$, set
$\beta=\norm{\R\a}_2$ and $k=k+1$.
\item (Enumeration at level $k$) Set $a_k=a_k+s_k$, $s_k=-s_k-$sgn$(s_k)$
and go to Step 3.
\end{enumerate}
\end{algorithm}

\subsection{Modified Schnorr-Euchner search algorithm}
\label{ss:Msearch}
In the following we make some comments to Algorithm \ref{a:search} and make some modifications.
It is easy to see that, the first nonzero integer vector encountered by Algorithm 2 is $\e_1$ and
the corresponding search radius is
\beq
\label{e:beta}
\beta=|r_{11}|=\sqrt{1-t_1^2}.
\eeq
Note that reordering the entries of $\bst$ that makes \eqref{e:tOrdered} hold gives the smallest $\beta$ among any other orderings.
This shows one of the benefits of the reordering leading to \eqref{e:tOrdered}. Also from \eqref{e:orderRnorm},
the reordering gives
$$
\beta=|r_{11}|=\min_{1\leq i\leq n}\norm{ \R_{1:i,i}}_2,
$$
which implies $\e_1$ is better than any other $\e_i$ for $i=2,\ldots,n$, as the former corresponds to the smallest residual.
In the modified algorithm, we just start with $\beta$ given by \eqref{e:beta}.

In Section \ref{sec:formR}, we mentioned that it is not necessary to form the entries of $\R$ explicitly; in the following,
we show how to compute $r_{kk}$ and $d_k$ for $1\leq k \leq n$, which are needed in \eqref{e:levelk}.  By \eqref{e:Ri} and \eqref{e:fi}, we have
\beq
\label{e:rkksquare}
r_{kk}^2=f_k/f_{k-1}, \quad 1\leq k\leq n.
\eeq
In the modified algorithm, we will use a $n$-dimensional vector $\q$ to store $r_{kk}^2$, i.e., let $q_{k}=r_{kk}^2$.

By \eqref{e:Ri}, \eqref{e:fi} and \eqref{e:d},
$$
d_k=\frac{t_k}{f_k}\sum_{j=k+1}^nt_ja_j.
$$
Thus, for computational efficiency, we introduce an $(n+1)$-dimensional vector
$\p$ with $p_{n+1}=0$ to store some computed quantities. Specifically, after $a_k,\,1\leq k\leq n$ is chosen in the search process,
we assume
\beq
\label{e:alphak}
p_k=p_{k+1}+t_ka_k, \quad 1\leq k\leq n,
\eeq
which explains why $p_{n+1}=0$. Therefore, we have
\beq
\label{e:dk}
d_k=\frac{t_kp_{k+1}}{f_k}, \quad 1\leq k\leq n.
\eeq

Now we make the main modification to Algorithm 2 by using the property of $\a^\star$ in \eqref{e:aOrdered}.
Note that in the search process of finding an integer point $\a$ in the hyper-ellipsoid,
the entries of $\a$ are determined in the following order: $a_n$, $a_{n-1}$, \ldots, $a_1$.
When we enumerate candidates for $a_n$ at level $n$, we will only enumerate the non-negative integers.
When we enumerate candidates for $a_k$ at level $k$ (note that at this point, $a_n$,
$a_{n-1}$, \ldots, $a_{k+1}$ have been chosen), we will only enumerate those greater than or equal to $a_{k+1}$.
By doing these we can prune a lot of nodes from the search tree to make the search process much faster.

For the users to implement the algorithm easily and for our later complexity analysis,
we provide the pseudo code of the modified algorithm in Algorithm \ref{a:Msearch}.

\begin{algorithm}
\setstretch{0.8}
\DontPrintSemicolon
\LinesNumbered
\SetAlgoCaptionSeparator{.}
\SetKwInOut{Input}{Input}
\SetKwInOut{Output}{Output}
\SetKwFunction{KwSign}{sign}
\SetKwFunction{KwAbs}{abs}
\SetKwFunction{KwSort}{sort}
\SetKwFunction{KwFloor}{floor}
\Input{A vector $\bst \in \Rbb^n$ that satisfies $\|\bst\|<1$ (see \eqref{e:t} and \eqref{e:tOrdered})}
\Output{A solution $\a^\star$ to \eqref{e:CFmodel}}

\BlankLine

$\f=\0^n, f_1=1-t_1^2$ \tcp*[h]{see \eqref{e:fi}}\;
$\q=\0^n, q_{1}=f_1$ \tcp*[h]{$q_k=r_{kk}^2$, see \eqref{e:rkksquare}}\;

\For{$i=2:n$}{
	$f_i=f_{i-1}-t_i^2$\tcp*[h]{see \eqref{e:fi}}\;
	$q_{i}=f_i/f_{i-1}$\tcp*[h]{\eqref{e:rkksquare}}\;
}

$\p=\0^{n+1}$ \tcp*[h]{see \eqref{e:alphak}}\;
$\d=\0^n$ \tcp*[h]{see \eqref{e:d}}\;
$\boldsymbol{\sigma}=\0^n$ \tcp*[h]{$\sigma_k\triangleq\sum_{i=k+1}^nr_{ii}^2(a_i-d_i)^2$ for $k<n$}\;
$k=1$\;
$\a=\e_1^{n+1}$ \tcp*[h]{intermediate solution}\;
$\a^\star=\e^n_1$\;
$\beta^2=q_1$ \tcp*[h]{see \eqref{e:fi} and \eqref{e:beta}}\;
$\delta=q_1$ \tcp*[h]{$\delta\triangleq q_{k}(a_k-d_k)^2$}\;
$\s=\textbf{1}^n$\;
$\mbox{flag}=\textbf{1}^n$ \tcp*[h]{flag variable}\;

\While{{\bf true}}{
	$\alpha = \sigma_k + \delta$\;

	\eIf{$\alpha<\beta^2$}{
		\eIf{$k>1$}{
			$p_{k}=p_{k+1}+t_ka_k$ \tcp*[h]{see \eqref{e:alphak}}\;
			$k=k-1$\;
			$\sigma_k = \alpha$\;
			$d_k=t_{k}p_{k+1}/f_{k}$ \tcp*[h]{see \eqref{e:dk}}\;
			$a_k=\lfloor d_k\rceil$\;
			$\mbox{flag}_k=0$\;
			
			\eIf{$a_k\leq a_{k+1}$}{
				$a_k=a_{k+1}$\;
				$\mbox{flag}_k=1$\;
				$s_k=1$\;
			}{
				$s_k=\mbox{sgn}(d_k-a_k)$ \tcp*[h]{see \eqref{e:sgn}}\;
			}
			
			$\delta=q_{k}(a_k-d_k)^2$\;
		}{
			$\beta^2=\alpha$\;
			$\a^\star=\a_{1:n}$\;
		}
	}{
		\eIf{$k<n$}{
			$k=k+1$\;
			$a_k=a_k+s_k$\;

			\uIf{$a_k=a_{k+1}$}{
				$\mbox{flag}_k=1$\;
				$s_k=-s_k-\mbox{sgn}(s_k)$\;
			}
			\uElseIf{$\mbox{\rm flag}_k=1$}{
				$s_k=1$\;
			}
			\Else{
				$s_k=-s_k-\mbox{sgn}(s_k)$\;
			}

			$\delta=q_{k}(a_k-d_k)^2$\;
		}{
			\Return\;
		}
	}
}

\caption{Finding the optimal coefficient vector based on sphere decoding }
\label{a:Msearch}
\end{algorithm}

Here we make a few comments to Algorithm \ref{a:Msearch}. To unify the enumeration strategies for level $n$ and for any lower level,
we set $\a$ to be an $(n+1)$-dimensional vector with $a_{n+1}\equiv0$, so that $a_k\geq a_{k+1}$ holds for $k=n$.
Since the optimal solution $\a^\star$ is $n$-dimensional, we save
$\a^\star=\a_{1:n}$ if $\a_{1:n}\neq\0$ when a valid integer vector $\a$ is found.
To avoid enumerating any integer smaller than $a_{k+1}$ at level $k$, we introduced a flag variable ``flag" in the
algorithm to indicate whether the enumeration has reached the lower bound $a_{k+1}$ for $1\leq k\leq n$.
In the algorithm $s_k$ is the difference between the next integer candidate for $a_k$ and the current value
of $a_k$ and it is used to get the next integer candidate for $a_k$.

\begin{remark}
By Theorem \ref{t:LBDs}, if \eqref{e:closdexp} holds, then $\e_1$ is the solution.
Therefore, before using the Modified Schnorr-Euchner search algorithm to find the optimal solution, we can test whether
\eqref{e:closdexp} holds. If it holds, then return $\e_1$, otherwise, we use the search algorithm to find the optimal solution.
This can usually further improve the efficiency of the algorithm, especially when $n$ or $P$ is small.
But from Table \ref{tb:optimal2}, we can see this case occurs in a very low probability when both $n$ and $P$ are very large.
Thus, for simplicity, we do not incorporate it in Algorithm \ref{a:Msearch}.
\end{remark}

\section{Complexity analysis}
\label{sec:Complexity}
In this section, we will analyze the complexity, in terms of flops, of the proposed method,
and compare it with  two optimal methods  proposed in \cite{RicSJ12} and \cite{SahG14},
and two suboptimal methods, which are the LLL reduction approach \cite{SakVHB12} and the quadratic programming
relaxation (QPR) approach \cite{ZhoM14} and its improvement in \cite{ZhoWM14}.
In the following analysis, we focus on the case that the channel entries are i.i.d. Gaussian,
i.e., $\h \sim \mathcal{N}\left(\0,\I\right)$.

\subsection{Complexity analysis for the modified Schnorr-Euchner search algorithm}
In this subsection, we analyze the complexity of Algorithm \ref{a:Msearch}.
The approach is to first estimate the number of nodes visited in the search tree and then to count the number of arithmetic operations for each node.

\begin{remark}
\label{r:FixedRadius}
It is difficult, if not impossible, to analyze the complexity of Algorithm  \ref{a:Msearch}
because  the search radius $\beta$ changes in the search process.
Thus, we assume that the search radius $\beta$ keeps unchanged in our following analysis
to obtain an upper estimate of the complexity of Algorithm  \ref{a:Msearch}.
\end{remark}

\begin{figure}[!htbp]
\centering
\includegraphics[width=160pt]{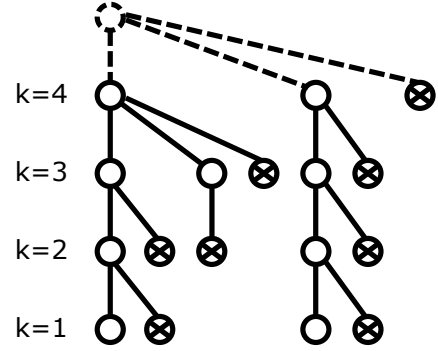}
\caption{An example search tree corresponding to Algorithm 2 with $\beta$ being a constant and $n=4$} \label{fig:tree}
\end{figure}

To illustrate our discussion,  we display the search tree corresponding to Algorithm  \ref{a:Msearch}
 with the assumption that $\beta$ is a constant in Figure \ref{fig:tree}.
Since there is not a true tree root, the dashed line is used for the root node  in Figure \ref{fig:tree}.
We will analyze the cost of this search tree, which is an upper bound on the complexity
of Algorithm  \ref{a:Msearch}.

To estimate the number of nodes at each level of the search tree, for $k=n,n-1,\ldots,1$, we define
\begin{align}
\label{e:Ek}
E_k(\beta)=\{\a_{k:n}\in\mathbb{Z}^{n-k+1}:\; &a_k\geq a_{k+1}\geq\ldots\geq a_n\geq 0,\nonumber\\
&\norm{\R_{k:n,k:n}\a_{k:n}}_2<\beta\}.
\end{align}
Note that each non-leaf node at level $k$ in the search tree corresponds to an $\a_{k:n} \in E_k(\beta)$,
and each leaf node labeled by $\times$ at level $k$ corresponds to the case that
$\a_{k:n}\notin E_{k}(\beta)$ with $\a_{k+1:n}\in E_{k+1}(\beta)$ ($k<n$).

Let $|E_k(\beta)|$ denote the number of elements belong to $E_k(\beta)$,
then the number of non-leaf nodes at level $k$ in the search tree  is $|E_k(\beta)|$.
It is easy to argue that the  number of  leaf nodes at level $k$ in the search tree is exactly equal to $|E_{k+1}(\beta)|$.
Thus the total number of nodes (including both the non-leaf and leaf nodes) at level $k$ is $|E_k(\beta)|+|E_{k+1}(\beta)|$.

From Algorithm \ref{a:Msearch}, any node at level $k$ ($k<n$) comes from two possibilities.
One is that it is generated after its parent node at level $k+1$ is generated.
This process corresponds to lines 19-35 of Algorithm \ref{a:Msearch} and the cost is $\bigO(1)$ flops.
The number of such nodes is $|E_{k+1}(\beta)|$. The other is that it is generated after a leaf node at level $k-1$ is generated.
This process corresponds to lines 38-49 and the cost is also $\bigO(1)$ flops. The number of such nodes is $|E_k(\beta)|$.
Thus, the total cost for generating all nodes at level $k$ is
\begin{eqnarray}
\label{e:costk}
c_k = (|E_{k}(\beta)|+|E_{k+1}(\beta)|)\bigO(1),
\end{eqnarray}
where we define $|E_{n+1}(\beta)|=0$.
Let $C(n)$ denote the total cost of the search tree,  then, by \eqref{e:costk}, we obtain
\beq \label{e:CnEk}
C(n) = \sum_{k=1}^n c_k=\bigO(1)\sum_{k=1}^n|E_{k}(\beta)|.
\eeq

Obviously, $|E_n(\beta)|\leq\lceil \beta/r_{nn}\rceil$. However, it is hard to rigorously
compute or estimate $|E_k(\beta)|$ for $k<n$ since the inequalities are involved in \eqref{e:Ek},
so for  $k=1,2,\ldots,n$, we define supersets:
\begin{align}
\label{e:Fk}
F_k(\beta)=\{\a_{k:n}\in\mathbb{Z}^{n-k+1}:\;\norm{\R_{k:n,k:n}\a_{k:n}}<\beta\},
\end{align}
where $\beta$ is the initial search radius used in Algorithm \ref{a:Msearch} (see \eqref{e:beta}).

Let $|F_k(\beta)|$ denote the number of elements belong to $F_k(\beta)$. Obviously, we have
\begin{align}
\label{ine:EFk}
|E_k(\beta)|\leq|F_k(\beta)|.
\end{align}
We apply the so-called Gaussian heuristic, which is widely used in the complexity analysis of sphere decoding methods
(see, e.g., \cite{GruW93, BanK98, AgrEVZ02,SeeJSB11}),
to estimate $|F_k(\beta)|$. This method approximates $|F_k(\beta)|$ by the volume of the hyper-ellipsoid $\norm{\R_{k:n,k:n}\a_{k:n}}_2<\beta$,
namely,
\beq
\label{e:nodeFapp}
|F_k(\beta)|\approx\frac{\beta^{n-k+1}}{\prod_{i=k}^nr_{ii}}V_{n-k+1},
\eeq
where $V_{n-k+1}$ denotes the volume of an $(n-k+1)$-dimensional unit Euclidean ball, i.e.,
\begin{eqnarray}
\label{e:volV}
V_{n-k+1}=\frac{\pi^{(n-k+1)/2}}{\Gamma((n-k+1)/2+1)}
\end{eqnarray}
with $\Gamma$ being the Gamma function.

By \eqref{e:Dk} and \eqref{e:beta},  we have
\begin{align}
\label{e:nodeF}
\frac{\beta^{n-k+1}}{\prod_{i=k}^nr_{ii}}&\leq \frac{ (1-\sum_{i=1}^{k-1}t_i^2)^{1/2}(1-t_1^2)^{(n-k+1)/2}}{\sqrt{1-\|\bst\|^2_2}}  \nonumber \\
& \leq  \frac{1}{\sqrt{1-\|\bst\|^2_2}}.
\end{align}

Since
\begin{equation*}
    \Gamma(\frac{n-k+1}{2}+1)=
   \begin{cases}
   (\frac{n-k+1}{2})!&\mbox{if $n-k$ is odd}\\
   \frac{\sqrt{\pi}(n-k+1)!!}{2^{(n-k)/2+1}} &\mbox{if $n-k$ is even}
   \end{cases},
\end{equation*}
where
$$
(n-k+1)!!=1\times3\times5\times\cdots \times (n-k+1).
$$
From \eqref{e:volV}, we have
\begin{equation}
\label{e:Vn-k+1}
   V_{n-k+1}=
   \begin{cases}
   \frac{\pi^{(n-k+1)/2}}{(\frac{n-k+1}{2})!}&\mbox{if $n-k$ is odd}\\
   \frac{2(2\pi)^{(n-k)/2}}{(n-k+1)!!} &\mbox{if $n-k$ is even}
   \end{cases}.
\end{equation}
By \eqref{e:Vn-k+1}, it is not hard to see that
$$
\lim_{k\mapsto \infty}V_k=0.
$$
Therefore, by \eqref{e:nodeF}, instead of using \eqref{e:nodeFapp}, we only need the following approximation,
which is weaker than the Gaussian heuristic,
\begin{align}
\label{e:nodeF2}
|F_k(\beta)|\lessapprox   \frac{1}{\sqrt{1-\|\bst\|^2_2}} \max\{V_{n-k+1},1\}.
\end{align}

By direct calculation, we have $V_{13}=0.9106, V_{14}=0.5993$.
By \eqref{e:Vn-k+1}, obviously, $V_i$ is decreasing with $i\geq 13$.
Therefore, from the aforementioned equation,
we obtain
\begin{equation*}
V_i\leq1, \;\, \forall \; \, i\geq 13.
\end{equation*}
By direct calculation, we have
\begin{equation*}
\max_{1\leq j\leq 12}V_j=V_5=5.2638.
\end{equation*}
Therefore, combining with \eqref{e:CnEk}, \eqref{ine:EFk} and \eqref{e:nodeF2}, we obtain (recall that $C(n)$ is the total cost of the search tree)
\begin{align}
\label{e:compbd1}
C(n)  \lessapprox \frac{\bigO(n)}{\sqrt{1-\|\bst\|^2_2}}.
\end{align}
By \eqref{e:t}, we have
\begin{align*}
\frac{1}{\sqrt{1-\|\bst\|^2_2}}=\frac{1}{\sqrt{1-\frac{P\|\h\|^2_2}{1+P\|\h\|^2_2}}}=\sqrt{1+P\|\h\|^2_2}.
\end{align*}
Thus, by \eqref{e:compbd1}, we obtain
\begin{align}
\label{e:compbd}
C(n)  \lessapprox \bigO(n)\sqrt{1+P\|\h\|^2_2}.
\end{align}

In the following, simulation results are provided to support that \eqref{e:compbd} holds for general $n$ and $\h$ in \eqref{e:t}.
From \eqref{e:CnEk}, we only need to show $\sum_{k=1}^n|E_{k}(\beta)| \lessapprox \bigO(n)\sqrt{1+P\|\h\|^2_2}$.
We consider the case that the channel vector $\h \sim \mathcal{N}\left(\0,\I\right)$.
For each $n$ and each $P$, we randomly generate $10000$ realizations of $\h$.

Table \ref{tb:ratio} displays the average and largest ratios of $\sum_{k=1}^n|E_{k}(\beta)|$ to $n\sqrt{1+P\|\h\|^2_2}$  over $10000$ samples.
``AR" and ``LR" in Table \ref{tb:ratio} respectively denote average and largest ratios.
From Table \ref{tb:ratio}, we can see that $\sum_{k=1}^n|E_{k}(\beta)|<2n\sqrt{1+P\|\h\|^2_2}$ in all the tests.
Note that the number of nodes searched by Algorithm \ref{a:Msearch} cannot be larger than $\sum_{k=1}^n|E_{k}(\beta)|$ because the radius $\beta$
reduces whenever a valid integer vector is found in the search process.
\begin{table}[tbp!]
\caption{Average and largest ratios (AR and LR) of $\sum_{k=1}^n|E_{k}(\beta)|$ to $n\sqrt{1+P\|\h\|^2_2}$  over $10000$ realizations of $\h$}
\scriptsize
\centering
\begin{tabular}{|c||c|c||c|c||c|c|}
 \hline
\backslashbox{$n$}{$P$}& \multicolumn{2}{c||}{$P=0\,\mbox{dB}$} &  \multicolumn{2}{c||}{$P=20\,\mbox{dB}$} & \multicolumn{2}{c|}{$P=40\,\mbox{dB}$} \\ \hline
& AR &  LR   &   AR &  LR & AR &  LR     \\ \hline
2  & 0.4241 & 1.4747 & 0.3032 & 1.3399 & 0.3178 & 0.9292 \\ \hline
4  & 0.5259 & 1.7803 & 0.4273 & 1.3123 & 0.4401 & 1.4314 \\ \hline
8  & 0.4408 & 1.2875 & 0.4040 & 1.2632 & 0.4253 & 1.5001 \\ \hline
16  & 0.3204 & 0.9109 & 0.1813 & 0.5917 & 0.1887 & 0.6826 \\ \hline
32  & 0.2205 & 0.6348 & 0.0610 & 0.1770 & 0.0509 & 0.2033 \\ \hline
64  & 0.1471 & 0.3783 & 0.0265 & 0.0602 & 0.0127 & 0.0394 \\ \hline
$10^2$  &  0.1143 & 0.2937 & 0.0183 & 0.0338 & 0.0054 & 0.0154 \\ \hline
$10^3$  &  0.0381 & 0.0567 & 0.0048 & 0.0063 & 0.0005 & 0.0006 \\ \hline
$10^4$  &  0.0129 & 0.0151 & 0.0014 & 0.0016 & 0.0001 & 0.0002 \\ \hline
$10^5$  &  0.0040 & 0.0043 & 0.0004 & 0.0004 & 0.0000 & 0.0000 \\ \hline
\end{tabular}
\label{tb:ratio}
\end{table}

In the following, we investigate the expected value of $C(n)$ when the entries of $\h$  are
independent and identically distributed following the normal distribution $\mathcal{N}(0,1)$.
It is easy to see that $\norm{\h}_2^2$ follows the Chi-squared distribution $\chi^2(n)$.
Therefore, $\mathbb{E}[\norm{\h}_2^2]=n$. Since $\sqrt{1+Px}$ is a concave function of $x$, by Jensen's Inequality,
\begin{align}
\label{e:JensenInequality}
\mathbb{E}\left[\sqrt{1+P\norm{\h}_2^2}\right]\leq
\sqrt{1+P\mathbb{E}\left[\norm{\h}_2^2\right]}
= \sqrt{1+nP}.
\end{align}
Therefore, by  \eqref{e:compbd} and \eqref{e:JensenInequality}, it is easy to see that the complexity of Algorithm \ref{a:Msearch}
is around $\bigO(n^{1.5})$ flops.

\subsection{Comparison of the complexity of the proposed method with other methods}

It is easy to see that, for any given $\h$, computing $\bst$ by \eqref{e:t} costs $\bigO(n)$ flops. And
for any fixed $\bst$, transform it such that \eqref{e:tOrdered} holds costs $\bigO(n\log(n))$ comparisons.
Since the total complexity of Algorithm \ref{a:Msearch} is around $\bigO(n^{1.5})$ flops,
the total complexity of the whole method is $\bigO(n^{1.5})$ flops for the test cases.

The complexity of the QPR in \cite{ZhoM14} and \cite{ZhoWM14} is $\bigO(n^{3})$ and $\bigO(n^{1.5})$ flops, respectively.
The method based on LLL lattice reduction \cite{SakVHB12} uses the regular method, costing $\bigO(n^3)$, to obtain the Cholesky factor $\R$.
The optimal method proposed in \cite{RicSJ12} needs to find the inverse of $n$ matrices and solving $n$ linear equations with the
dimensions from 1 to $n$, so its complexity is not smaller than $\bigO(n^{3})$.
The complexity of the optimal method proposed in \cite{SahG14} is $\bigO(n^{2.5})$ flops.
Therefore, it is expected that our optimal algorithm is faster than the LLL reduction based method,
the QPR  in \cite{ZhoM14} and the two optimal methods proposed in \cite{SakVHB12} and \cite{SahG14},
and is faster than or has more or less the same speed as the QPR  in \cite{ZhoWM14}.

\section{Numerical simulations}
\label{sec:simcf}
In this section, we present the numerical results to illustrate the effectiveness and efficiency
of our new method. We consider the case that the entries of the channel vector $\h \in \mathbb{R}^n$ are i.i.d. Gaussian,
i.e., $\h \sim \mathcal{N}\left(\0,\I\right)$.
The dimension $n$ of $\h$ ranges from 2 to 16.
For a given $n$, we randomly generate $10000$ realizations of $\h$ for each $P$ from $0$ dB to $20$ dB,
and apply different methods to calculate the corresponding computation rates.
To compare the effectiveness of different methods, we compute the average computation rates. To compare their efficiency, we record the running time.

The methods considered include our new method called the improved sphere decoding (ISD) method,
the branch-and-bound (BnB) algorithm in \cite{RicSJ12}, the optimal method proposed in \cite{SahG14} (abbreviated as SG named after the authors),
the method based on LLL lattice reduction algorithm \cite{SakVHB12} (abbreviated as LLL),
and the quadratic programming relaxation (QPR) approach \cite{ZhoWM14}.
The \emph{quality-complexity tradeoff factor} $\delta$
in the LLL method is set as $0.75$. A larger $\delta$ ($1/4<\delta\leq1$) can give a higher rate, but the running
time will increase drastically as $\delta$ increases.
\emph{The upper bound on the number of real-valued approximations}, $K_u$,
in the QPR method is set according to the criterion proposed in \cite{ZhoWM14}.
Exact values of $K_u$ used in the simulations are listed in Table~\ref{tab:Ku}.

\begin{figure}[!ht]
    \centering
    \subfloat[$n=4$]{
        \label{fig:Rate_n4}
        \includegraphics[width=220pt]{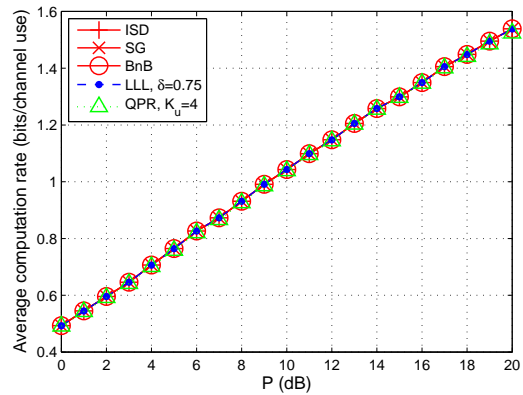}}\\
    \subfloat[$n=8$]{
        \label{fig:Rate_n8}
        \includegraphics[width=220pt]{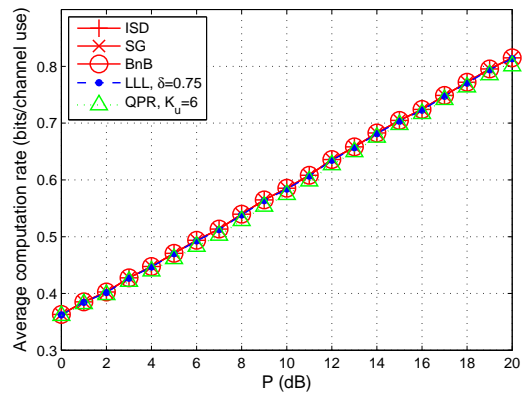}}\\
    \subfloat[$n=16$]{
        \label{fig:Rate_n16}
        \includegraphics[width=220pt]{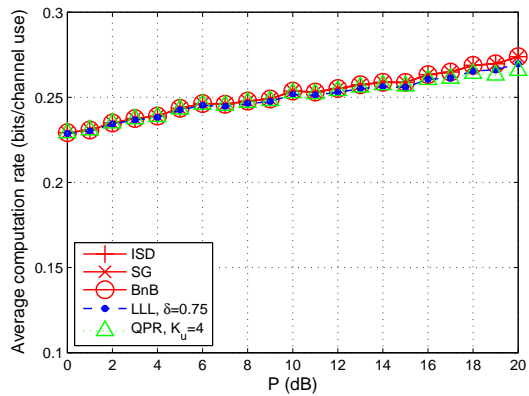}}
    \caption{Average computation rates by different methods.}
    \label{fig:RateComparison}
\end{figure}

\setlength{\tabcolsep}{3pt} 
\begin{table}[h!]
  \caption{The upper bound on the number of real-valued approximations in QPR method}
  \label{tab:Ku}
  \centering
  \begin{tabular}{c|*{15}{r}}
  \hline
  $n$ &2 & 3 & 4 & 5 & 6 & 7 & 8 & 9 & 10 & 11 & 12 & 13 & 14 & 15 & 16\\
  \hline
  $K_u$ &2 & 3 & 4 & 5 & 5 & 5 & 6 & 6 & 6 & 6 & 7 & 6 & 6 & 6 & 4\\
  \hline
  \end{tabular}
\end{table}

We first compare the average computation rates.
Figure~\ref{fig:RateComparison}(a),~\ref{fig:RateComparison}(b),~\ref{fig:RateComparison}(c)
show the average computation rates over 10000 samples
with the dimension $n$ being 4, 8, and 16, respectively.
The ISD method, the BnB method and the SG methods are optimal.
As expected, numerical results show that they always  provide the highest computation rate.
The corresponding curves of these three methods in Figure~\ref{fig:RateComparison}
exactly overlap with each other.
The QPR method and the LLL based method provide rates close to that of the optimal methods.
However, as the dimension increases, their performance degrade.

\begin{figure}[!htbp]
    \centering
    \subfloat[$P=0$ dB]{
        \label{fig:Time_PdB0_delta075}
        \includegraphics[width=220pt]{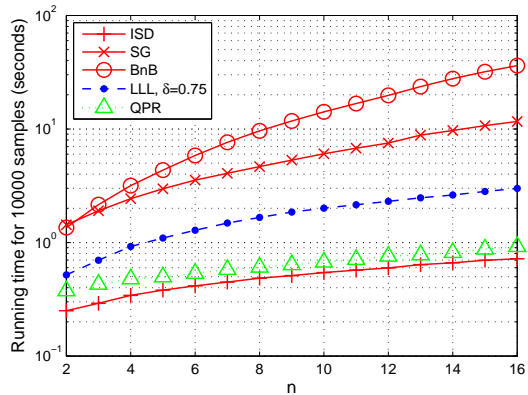}}\\
    \subfloat[$P=10$ dB]{
        \label{fig:Time_PdB10_delta075}
        \includegraphics[width=220pt]{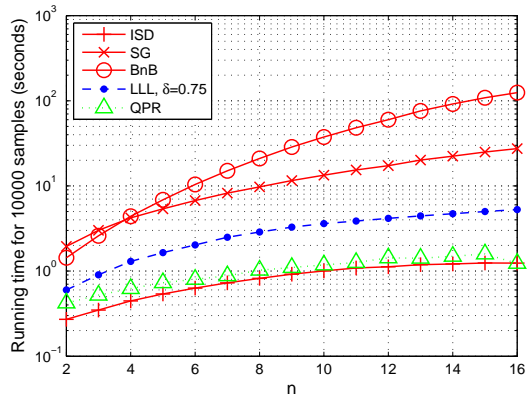}}\\
    \subfloat[$P=20$ dB]{
        \label{fig:Time_PdB20_delta075}
        \includegraphics[width=220pt]{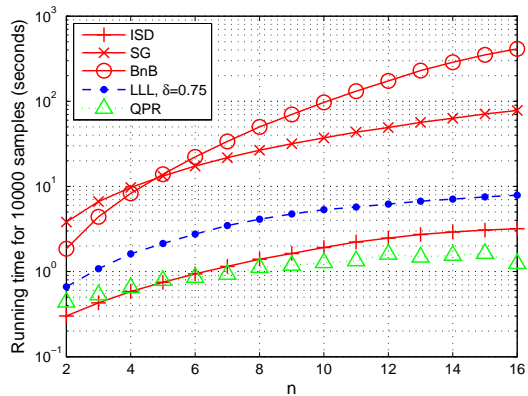}}
    \caption{Running time of different methods for $10000$ samples .}
    \label{fig:TimeComparison}
\end{figure}

Now we compare the running time.
Figure~\ref{fig:TimeComparison}(a),~\ref{fig:TimeComparison}(b),~\ref{fig:TimeComparison}(c)
show the running time of simulating $10000$ samples with $P$ being $0$ dB, $10$ dB, and $20$ dB, respectively.
For the optimal methods, it is obvious that our new ISD method is much more efficient than both the BnB method and SG method.
It can also be observed that the ISD method is also faster than the LLL based method.
Although the QPR method \cite{ZhoWM14} is faster than our ISD method in Figure~\ref{fig:TimeComparison}(c), it is a suboptimal method and
its performance degrades for high dimension (see Figure~\ref{fig:RateComparison}(c)).

\section{Adaptation for Use in Compute-and-Forward Design}
\label{sec:Adaptation}
In the previous sections, we mainly focused on the problem of finding
\emph{the optimal coefficient vector for one relay},
which is defined as the one that gives the highest computation rate at that relay.
However, it is the overall \emph{transmission rate} that
determines the system performance, rather than the computation rate at a single relay.
The coefficient vector that maximizes the computation rate at one relay may not be
optimal for the whole system.
Thus, in this section we show how to modify our algorithm so that
it can be applied in the CF design.
We will also analyze the complexity of the modified algorithm,
and present numerical results of the running time
to show the efficiency of the modified algorithm.

\subsection{Adapting the proposed algorithm}
In this subsection, we show how to adapt our algorithm for use in CF design.

The \emph{transmission rate} is the minimum computation rate over all relays
if the coefficient matrix,
which is formed by the coefficient vectors at relays, is full column rank;
and it is 0 if the coefficient matrix has rank deficiency~\cite{NazG11}.
One naive strategy is to let each relay choose the coefficient vector that
maximizes the computation rate. However, it has been shown in~\cite{WeiC12} to be
inherently suboptimal
since there is a high probability that the formed coefficient
matrix is not full rank, especially for low SNR.
Instead, we can adopt the strategy (named ``Wei-Chen" after the authors)
proposed in~\cite{WeiC12}:
\begin{enumerate}
\item Each relay searches a list of candidate coefficient vectors,
and then forwards the list to the destination.
\item The destination performs a search based on the received lists,
and finds a good set of coefficient vectors that can form a full rank coefficient matrix,
and then sends to each relay the coefficient vector be used.
\item Each relay chooses the coefficient vector to be the one it receives.
\end{enumerate}
The Wei-Chen strategy effectively resolves the rank deficiency issue,
and achieves close-to-optimal transmission rate.

\begin{algorithm}
\setstretch{0.8}
\DontPrintSemicolon
\LinesNumbered
\SetAlgoCaptionSeparator{.}
\SetKwInOut{Input}{Input}
\SetKwInOut{Output}{Output}
\SetKwFunction{KwSign}{sign}
\SetKwFunction{KwAbs}{abs}
\SetKwFunction{KwSort}{sort}
\SetKwFunction{KwFloor}{floor}
\Input{A vector $\bst \in \Rbb^n$ that satisfies $\|\bst\|<1$ (see \eqref{e:t} and \eqref{e:tOrdered}),\\
and the desired number $L$ of candidates.}
\Output{A list $\Omega$ containing $L$ (or less, see Remark~\ref{r:ListLength}) best integer vectors
to \eqref{e:CFmodel},\\
and a list $\Gamma$ of the corresponding objective values $\norm{\R\a}$ for $\a\in\Omega$.}

\BlankLine

$\f=\0^n, f_1=1-t_1^2$ \tcp*[h]{see \eqref{e:fi}}\;
$\q=\0^n, q_{1}=f_1$ \tcp*[h]{$q_k=r_{kk}^2$, see \eqref{e:rkksquare}}\;

\For{$i=2:n$}{
	$f_i=f_{i-1}-t_i^2$\tcp*[h]{see \eqref{e:fi}}\;
	$q_{i}=f_i/f_{i-1}$\tcp*[h]{\eqref{e:rkksquare}}\;
}

$\p=\0^{n+1}$ \tcp*[h]{see \eqref{e:alphak}}\;
$\d=\0^n$ \tcp*[h]{see \eqref{e:d}}\;
$\boldsymbol{\sigma}=\0^n$ \tcp*[h]{$\sigma_k\triangleq\sum_{i=k+1}^nr_{ii}^2(a_i-d_i)^2$ for $k<n$}\;

$\Omega=\O$\;
$\Gamma=\O$\;
$k=1$\;
$\a=\e^n_1$ \tcp*[h]{intermediate solution}\;
$\beta^2=1$ \tcp*[h]{see \eqref{ine:ellp}}\;
$\delta=q_1$ \tcp*[h]{$\delta\triangleq q_{k}(a_k-d_k)^2$}\;
$\s=\textbf{1}^n$\;

\While{{\bf true}}{
	$\alpha = \sigma_k + \delta$\;

	\eIf{$\alpha<\beta^2$}{
		\eIf{$k>1$}{
			$p_{k}=p_{k+1}+t_ka_k$ \tcp*[h]{see \eqref{e:alphak}}\;
			$k=k-1$\;
			$\sigma_k = \alpha$\;
			$d_k=t_{k}p_{k+1}/f_{k}$ \tcp*[h]{see \eqref{e:dk}}\;
			
			\eIf{$k>1$}{
				$a_k=\round{d_k}$\;
				$s_k=\mbox{sgn}(d_k-a_k)$ \tcp*[h]{see \eqref{e:sgn}}\;
			}{
				$a_1=\ceil{d_1}$\;
				$s_1=1$\;
			}
			
			$\delta=q_{k}(a_k-d_k)^2$\;
		}{
			\eIf{$\card{\Omega}=L$}{
				$m=\arg\max_i \gamma_i$ \tcp*[h]{$\Gamma=\{\gamma_i\}$}\;
				$\omega_m=\a$ \tcp*[h]{$\Omega=\{\omega_i\}$}\;
				$\gamma_m=\alpha$\;
			}{
				$\Omega=\{\Omega, \a\}$\;
				$\Gamma=\{\Gamma, \alpha\}$\;
			}
            \If{$\card{\Omega}=L$}{
				$\beta^2=\max_i \gamma_i$\;
			}
			$a_1=a_1+s_1$\;
			$\delta=q_{1}(a_1-d_1)^2$\;
		}
	}{
		\eIf{$k<n$}{
			$k=k+1$\;
			$a_k=a_k+s_k$\;
			$s_k=-s_k-\mbox{sgn}(s_k)$\;
			$\delta=q_{k}(a_k-d_k)^2$\;
		}{
			\Return\;
		}
	}
}

\caption{Finding $L$ best coefficient vectors based on sphere decoding}
\label{a:MSearchList}
\end{algorithm}

To apply our algorithm along with the Wei-Chen strategy,
we need to modify our search algorithm such that
it outputs \emph{a list of best coefficient vectors providing the best rates}.
In fact, several slight modifications suffice to serve the purpose:
\begin{enumerate}
\item Discard the constraint on the candidates in~\eqref{e:aOrdered}.
This is to include suboptimal candidates that do not satisfy~\eqref{e:aOrdered}
but provide close-to-optimal rates.
\item Enumerate vectors at level-1 in the natural order by setting
the initial value as $a_1=\ceil{d_1}$ in~\eqref{e:d} and the step fixed as $s_1=1$.
Since $\a$ and $-\a$ result in the same rate,
only one of them needs to be enumerated.
The above way of enumeration at level 1 ensures that only one of $\a$ and $-\a$ is enumerated.
\item Initialize the radius $\beta$ in~\eqref{ine:ellp} as 1 and shrink $\beta$ appropriately
when a new candidate is enumerated:
if the number of candidates in the output list is less than the desired number,
put the new candidate in the list, and do not shrink the radius;
otherwise, replace the candidate who has the largest value of $\norm{\R\a}$ in~\eqref{ine:ellp}
(the most suboptimal candidate) in the the list with the new candidate,
and shrink the radius as the corresponding radius of the most suboptimal candidate in the updated list.
It is sufficient to set the initial radius as 1 rather than a larger value,
because coefficient vectors providing positive rates must have radius smaller than 1.
\end{enumerate}
The pseudo-code of the algorithm with the above modifications is provided
in Algorithm~\ref{a:MSearchList}.

\begin{remark}
\label{r:ListLength}
Note that Algorithm~\ref{a:MSearchList} may output a list containing less candidates than desired.
The reason is there are cases that the number of candidates that give positive rates
is less than the desired number, and it is meaningless to contain vectors with 0 rate in the list.
\end{remark}

\begin{remark}
\label{r:OutputTransform}
The integer vectors in the output list of Algorithm~\ref{a:MSearchList} need to be transformed
as described in Section~\ref{a:Reordering} to serve as the CF coefficient vectors.
\end{remark}

\subsection{Complexity analysis and numerical results}
In this subsection, we will first analyze the complexity of our modified search algorithm,
and then give numerical results of running time to show its efficiency.

The modifications that transform Algorithm~\ref{a:Msearch} with a single output
to Algorithm~\ref{a:MSearchList} with list output increase the running time for every instance.
However, the upper bound $\bigO(n)\sqrt{1+P\|\h\|^2_2}$ on the estimated number of tree nodes searched
by Algorithm~\ref{a:Msearch} still applies to Algorithm~\ref{a:MSearchList},
since in the complexity analysis presented in Section~\ref{sec:Complexity}
we have already assumed the radius $\beta$ does not shrink during the search (see Remark~\ref{r:FixedRadius}).
In Algorithm~\ref{a:MSearchList}, updating the length-$L$ output list
after a new candidate with better objective is found takes $\bigO\big(n+\log(L)\big)$ flops:
$\bigO\big(\log(L)\big)$ flops for locating the entry to be updated,
and $\bigO(n)$ flops for updating the located entry by replacing the entry with the new candidate.
Outputting the length-$L$ list takes $\bigO(nL)$ flops.
Thus, the average complexity of Algorithm~\ref{a:MSearchList} is estimated to be
$\bigO\left(n\big(n+\log(L)\big)\right)\sqrt{1+P\|\h\|^2_2} + \bigO(nL)$.
For the channel model we consider where $\h$ has entries being i.i.d. standard Gaussian,
the estimated average complexity of Algorithm~\ref{a:MSearchList} is
$\bigO\left(n^{2.5} + n^{1.5}\log\big(L\big) + nL\right)$.

\begin{figure}[!htb]
    \centering
    \subfloat[$P=10$dB]{
        \label{fig:ListSDvsFP_Time_PdB10}
        \includegraphics[width=220pt]{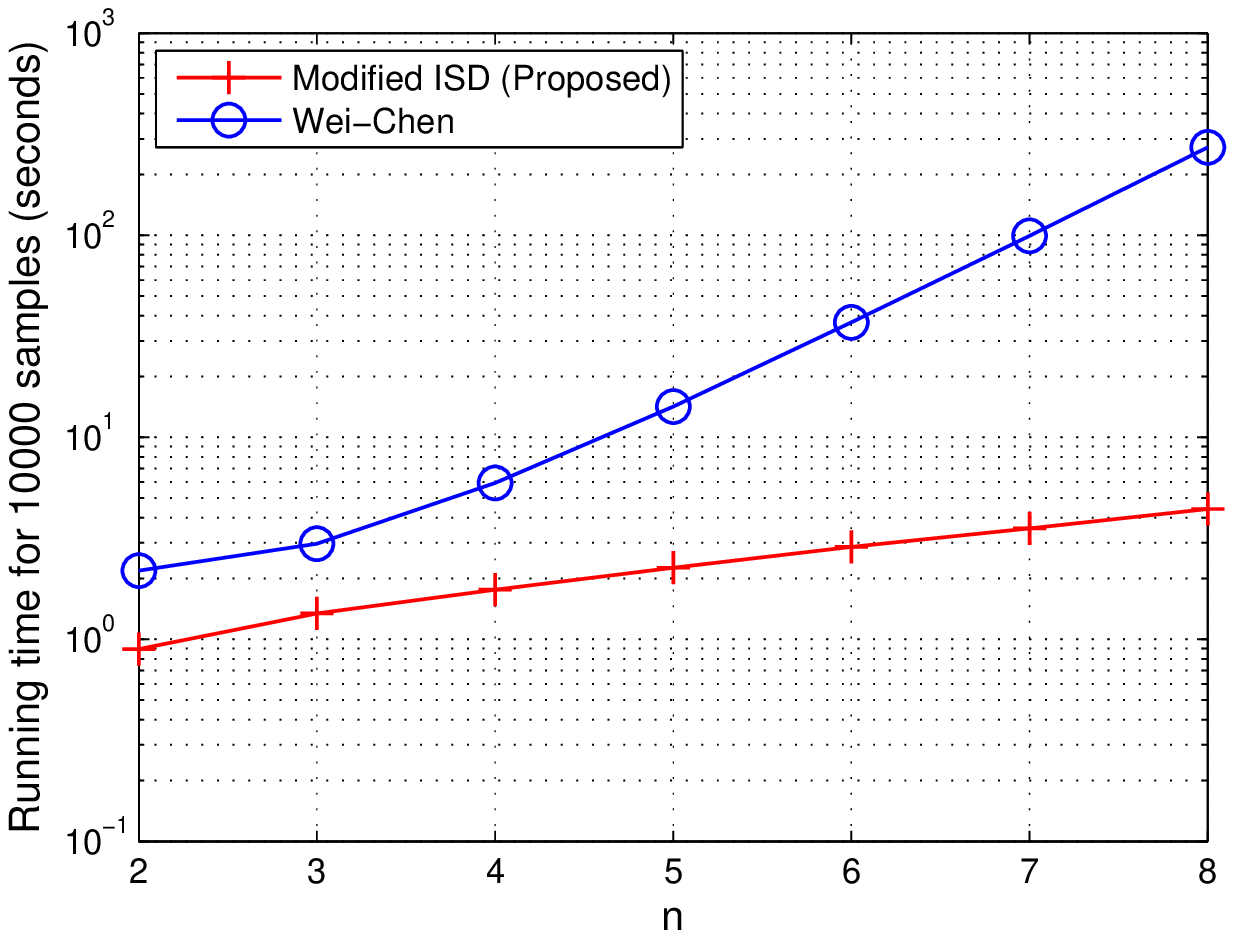}}\\
    \subfloat[$n=8$]{
        \label{fig:ListSDvsFP_Time_n8}
        \includegraphics[width=220pt]{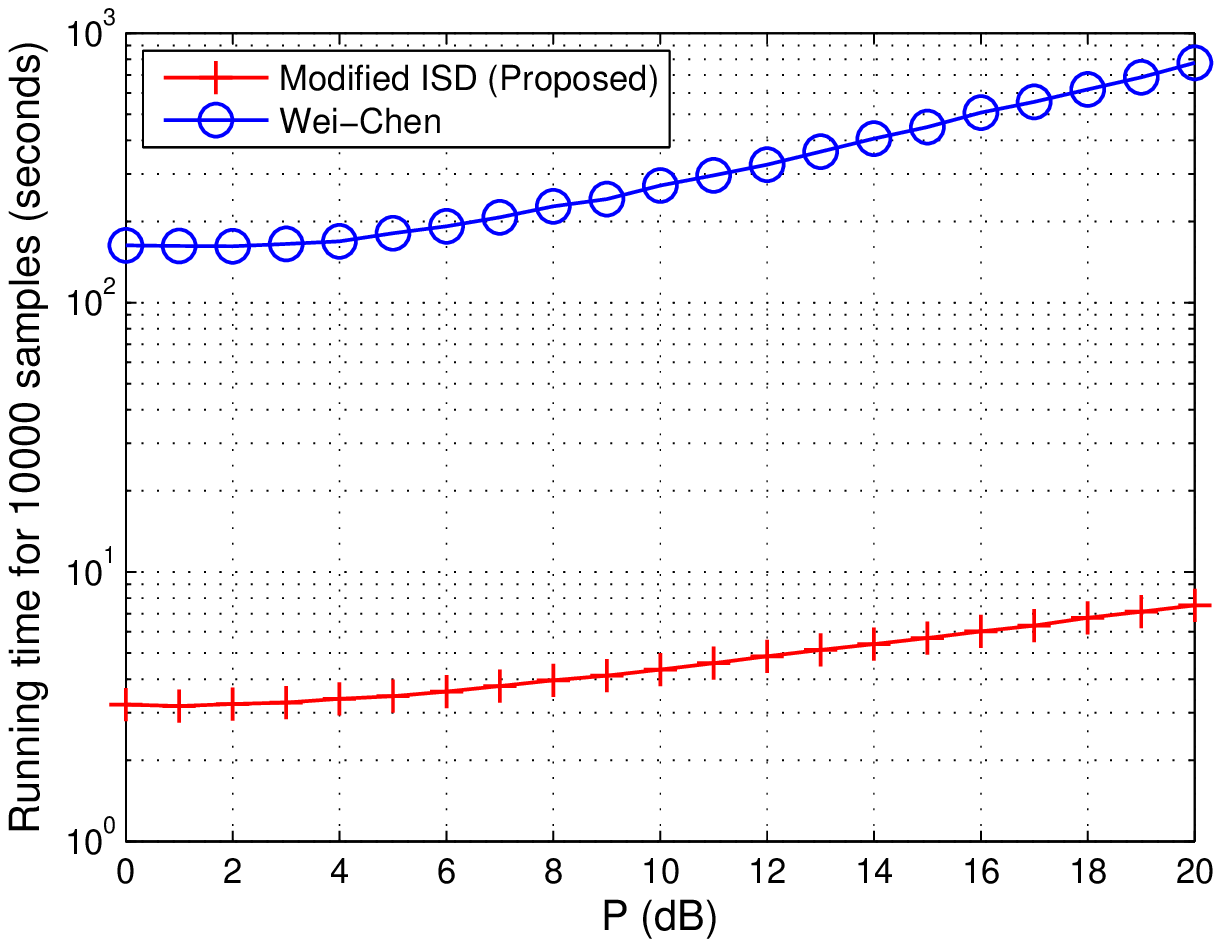}}
    \caption{Running time for 10000 samples by different methods with output list size $L=5$.}
    \label{fig:ListSDvsFP}
\end{figure}

Now we present numerical results of the running time to show the efficiency of our algorithm.
We consider the wireless relay network with $n$ sources and $n$ relays.
Slow fading channel with entries being i.i.d. standard Gaussian is assumed as before.
For each pair of $n$ and $P$ shown in the figures below,
10000 instances of the channel vector $\h$ are randomly generated.
In Figure~\ref{fig:ListSDvsFP_Time_PdB10} and \ref{fig:ListSDvsFP_Time_n8},
``Modified ISD" refers to our Algorithm~\ref{a:MSearchList} with list output;
while for comparison purpose, ``Wei-Chen" is the algorithm proposed by Wei and Chen in~\cite{WeiC12},
which is based on the Fincke-Pohst search.
When recording the running time, the time of transforming the output of Algorithm~\ref{a:MSearchList}
as stated in Remark~\ref{r:OutputTransform} is also included.
The two algorithms output the same candidate list except for the cases stated in Remark~\ref{r:ListLength},
where ``Wei-Chen" algorithm outputs additional vectors with 0 rate
so that the output list is of the given length $L$.

Figure~\ref{fig:ListSDvsFP_Time_PdB10} shows the running time in seconds
of the two considered algorithms with SNR $P=10$dB for different dimension $n$.
It is clear that our algorithm is much more efficient than the other.
The improvement in efficiency grows dramatically as the dimension $n$ increases:
at $n=2$ our algorithm is more than 2 times faster than the other algorithm,
while at $n=8$ our algorithm is more than 60 times faster,
and the running time saving goes beyond 98\%!

Figure~\ref{fig:ListSDvsFP_Time_n8} shows the running time in seconds
of the two considered algorithms with dimension $n=8$ for different SNR $P$.
As can be seen, our algorithm is consistently much more efficient than the other algorithm,
and the improvement is universally significant for SNR from as low as 0dB to as high as 20dB.

\section{Conclusions}
\label{sec:Conclusions}

Based on the idea of sphere decoding, in this paper, a new low-complexity algorithm, which
gives the optimal coefficient vector that maximizes the computation rate for a relay in the compute-and-forward scheme is proposed.
We derived an efficient algorithm to compute the Cholesky factorization by using the special structure of the Gram matrix.
It transformed the  problem into a SVP in $\bigO(n)$ flops without explicitly forming the whole Cholesky factor matrix.
Some conditions, under which $\e_1$ is an optimal coefficient vector,
have also been given, and can be checked in $\bigO(n)$ flops.
We then modified the Schnorr-Euchner search algorithm to solve the SVP by taking advantage of the properties of the optimal coefficient vector.
We showed that the expected complexity of our new method is $\bigO(n^{1.5})$ for i.i.d. Gaussian channel entries based on the Gaussian heuristic.
Simulations showed that our optimal method is not only much more efficient
than the existing ones that give the optimal computation rate,
but is also more efficient than some best previously known methods that give the close-to-optimal rate.
In addition, we demonstrated how to adapt our algorithm so that it can be applied
in compute-and-forward design.

\section*{Acknowledgment}
Part of this work was undergone while the first author was
visiting Hong Kong University of Science and Technology,
and studying as a Ph.D student at McGill University.
The hospitality received in the above periods is gratefully acknowledged.

\bibliographystyle{IEEEtran}
\bibliography{ref}

\end{document}